\documentclass[12pt]{iopart}
\usepackage{graphicx}
\usepackage{dcolumn}
\usepackage{bm}
\usepackage{epsf}
\usepackage{float}
\usepackage{iopams}

\def\ast{{\dagger}}

\def\Tc0{  { T_{c}^{0}    }  }
\def\INT01{  {   \dsint_{0}^{1} dq_1 dq_2 dq_3    }  }
\def\tiltc0{\tilde T_{c}^{0}}

\newcommand{\edc}{\end{document}}
\newcommand{\bb} {\begin{thebibliography}}
\newcommand{\eb}{\end{thebibliography}}
\newcommand{\bc}{\begin{center}}
\newcommand{\ec}{\end{center}}

\newcommand{\be}{\begin{equation}}
\newcommand{\ee}{\end{equation}}
\newcommand{\bea}{\begin{eqnarray}}
\newcommand{\eea}{\end{eqnarray}}
\newcommand{\ba}{\begin{array}{l}   }
\newcommand{\lab}[1]{\label{#1}}
\newcommand{\ea}{\end{array}}
\newcommand{\dsfrac}{\displaystyle\frac}
\newcommand{\ds} {\displaystyle}

\newcommand{\re}[1]{(\ref{#1})}

\newcommand{\ci}{\cite}

\newcommand{\dsint}{\ds\int}

\def\bfr{{\bf r}}
\def\bfk{{\bf k}}

\newcommand{\veps}{\varepsilon }

\begin{document}
\title[Bose- Einstein condensation of triplons]{Bose- Einstein condensation of triplons with a weakly broken U(1) symmetry}
\author{Asliddin Khudoyberdiev $^{1}$, Abdulla Rakhimov$^{2}$ and
Andreas Schilling  $^{3}$ }
\address{
$^1$ Institute of Nuclear Physics, Tashkent, 100214, Uzbekistan}
\address{$^2$ National University of Uzbekistan, Tashkent 100174, Uzbekistan}
\address{$^3$ Physik-Institut, University of Z\"{u}rich, Winterthurerstrasse 190, 8057 Z\"{u}rich, Switzerland}
\ead{asliddinkh@gmail.com}
\ead{rakhimovabd@yandex.ru}
\ead{schilling@physik.uzh.ch}

\begin{abstract}

The low-temperature properties of certain quantum magnets can be described in terms of a Bose-Einstein condensation (BEC) of magnetic quasiparticles (triplons). Some mean-field approaches (MFA) to describe these systems, based
on the standard grand canonical ensemble,
 do not take the anomalous density into account and leads to an internal inconsistency, as it has been shown by Hohenberg and Martin, and may therefore produce unphysical results. Moreover, an explicit breaking of the U(1) symmetry as observed, for example, in TlCuCl$_3$ makes the application of MFA more complicated. In the present work, we develop a self-consistent MFA approach, similar to the Hartree-Fock-Bogolyubov  approximation in the notion of representative statistical ensembles, including the effect of a weakly broken U(1) symmetry. We apply our results on experimental data of the quantum magnet TlCuCl$_3$ and show that magnetization curves and the energy dispersion can be well described within this approximation assuming that the BEC scenario is still valid. We predict that the shift of the critical temperature $T_c$ due to a finite exchange anisotropy is rather substantial even when the anisotropy parameter $\gamma$ is small, e.g.,
$\Delta T_c \approx 10 \%$ of $T_c$   in $H $= 6 \rm{T} and for $\gamma\approx 4 \rm{\mu eV}$.
\end{abstract}
\submitto{njp}
\maketitle

\section{Introduction}

Spontaneous symmetry breaking (SSB) plays an important role in particle and condensed matter physics. In the  Standard Model of particle physics
  SSB of gauge symmetries
 is responsible for generating masses for several particles and separating the electromagnetic and weak forces \ci{cheng}. In condensed matter physics SSB lies in the origin of effects such as Bose-Einstein condensation (BEC), superconductivity, ferromagnetism etc. In terms of microscopical field-theoretical models SSB corresponds to the case when the Hamiltonian of the system is invariant under a given transformation, while its ground state is not. Particularly, BEC is related to the U(1) symmetry of the Hamiltonian as $\psi(\mathbf{r})\rightarrow \psi(\mathbf{r})e^{i\alpha}$ for the field operator $\psi(\mathbf{r})$, with $\alpha$ being a real number.
 Moreover, strictly speaking, spontaneous symmetry breaking is a sufficient condition for the occurrence of BEC \ci{ginibre}.
Remarkably, not only real particles, but also quasiparticles may undergo BEC. In 1999 Oosawa \textit{et al} \ci{oosawa} performed magnetization measurements to investigate the critical behavior of the field-induced  magnetic ordering in the quantum antiferromagnetic TlCuCl$_{3}$. Changing the external magnetic field in the range of $H_{\rm{ext}} \sim$ 5 \rm{T} $\div$ 7 \rm{T} they observed an unexpected inflection of the  magnetization curve, i.e. $M(T,H)$, when $H_{\rm{ext}}$ exceeds a critical value, $H_{\rm{ext}}>$ $H_c$.

In fact, as it is seen from Fig.3 of Ref. \ci{oosawa} , for $H_{\rm{ext}}>$5.3 \rm{T} there is a critical temperature $T_{c}$($H_{\rm{ext}}$) below which the magnetization of the antiferromagnet starts to increase.
Later on due to the works by  R\"{u}egg  \ci{rueggnat},  Yamada  \ci{yamada} and
Nikuni  \ci{nikinu}, who obtained similar results as in  \ci{oosawa}, the  following interpretation of this phenomenon has been established :
{\begin{enumerate}
\item
In some compounds such as KCuCl$_{3}$ or TlCuCl$_{3}$ two Cu$^{2++}$ ions are antiferromagnetially coupled to form a dimer in a crystalline network: the dimer ground state is a spin singlet (S=0), separated by an energy gap from the first excited triplet state with S=1.
\item
At a critical external magnetic field, the energy of one of the Zeeman split triplet components intersects the ground state singlet and the gap between these two state may be closed.
\item
The appropriate quasiparticles, in the following called triplons, undergo BEC below a critical temperature, $T\leq T_{c}$.
\item
The whole density of triplons, $\rho$ and the density of condensed triplons, $\rho_{0}$, defines the  $M_{\parallel}$ and $M_{\perp}$ magnetizations per Cu atom. Namely,  $M_{\parallel}\sim\rho$ and $M_{\perp}\sim\sqrt{\rho_{0}}$.
\item
The density of triplons is directly controlled by the applied magnetic field
which acts as a chemical potential.
\item
 The thermodynamic characteristics as well as the magnetization may be calculated with a simple effective  Hamiltonian
 \begin{equation}
H=\sum_{\mathbf{k}}(\varepsilon_{k}-\mu)a_{\mathbf{k}}^{\dag}a_{\mathbf{k}}+\frac{U}{2}\sum_{\mathbf{k},\mathbf{k}',\mathbf{q}}a_{\mathbf{k}+\mathbf{q}}^{\dag}a_{\mathbf{k}'-\mathbf{q}}^{\dag}a_{\mathbf{k}}a_{\mathbf{k}'}
\end{equation}
where $\varepsilon_{\mathbf{k}}$ is the kinetic energy determined by  the dispersion around the lowest excitation, $\mu$ is the chemical potential given by
\begin{equation}
\mu=g\mu_{B}(H_{\rm{ext}}-H_{c}),
\lab{much1}
\end{equation}
$U$ is the interaction constant, and $a_{\mathbf{k}}^{\dag}(a_{\mathbf{k}})$ are creation (annihilation) operators for a triplon with momentum $\mathbf{k}$.
 \end{enumerate}
Further  similar effects with triplon condensation
\footnote{The difference between magnons and triplons and their possible condensation is
discussed in Ref. \ci{yuktriplon} }
 have been observed in  other quantum magnets and have been reviewed in \ci{zapf}.

Now we note that, besides of SSB, there is one more necessary condition for the existence of a  condensate. It concerns the  spectrum of collective excitations $E_{k}$ and is related to the Goldstone theorem. This condition reads
\begin{equation}  \label{limit}
\lim_{\mathbf{k}\rightarrow \mathbf{k}_{0}} E_{k}\sim c |\mathbf{k}-\mathbf{k}_0|,
\end{equation}
where $\mathbf{k}_{0}$ is a microscopically occupied single state and $c$ is the sound velocity.
The condition (\ref{limit}) along with stability conditions,
$\rm{Re (E_{k})\geq 0}$, $\rm{Im (E_{k})\leq 0}$ means that the collective excitation of the BEC state should be gapless as was observed by R\"{u}egg \textit{et al.} \ci{rueggnat} by neutron  scattering measurements  within their experimental resolution. Thus one arrives  at the  preliminary conclusion that
 experiments on
 magnetization and excitation energy made on TlCuCl$_{3}$ may be well described in terms of BEC of triplons \ci{rueggnat}-\ci{zapf}.

However, electron spin resonance (ESR) \ci{cizmar} and inelastic neutron scattering (INS)
 \ci{rueggprl} experiments on quantum antiferromagnets show an anisotropy of the spectrum of magnetic excitations which means that the corresponding O(3) (or equivalently U(1) symmetry in terms of bosons) in the plane perpendicular to the magnetic field is broken. The degree of explicit U(1) symmetry breaking is negligibly small for some materials (e.g. $\Delta U\sim 0.7$ \rm{mK} for BaCuSi$_2$O$_6$) and rather large for others (e.g. $\Delta U\sim 0.28$ \rm{K} for TlCuCl$_3$)
\footnote{see Table 1 in ref. \ci{zapf}}. Clearly, uniaxially symmetry breaking may be caused
in real quantum magnets by the effective spin- spin interactions induced by spin - orbit coupling or dipole -dipole interactions.

The presence of anisotropies violating  rotational symmetry in real magnetic materials may modify the physics, especially in the vicinity of the quantum critical points
\ci{giamarchi}. Particularly, because of explicit breaking of U(1) symmetry the BEC - scenario does not work, and hence there is no Goldstone mode because the energy spectrum acquires  a gap. Moreover, in the ESR measurements \ci{cizmar} a direct singlet-triplet transition has been observed which means that the gap cannot be completely closed with the  Zeeman effect. This mixing of the singlet and triplet states suggests that one must include an additional term into the Hamiltonian such as
\begin{equation} \label{hamil}
H_{\rm{DM}}'=i\gamma'\sum_{\mathbf{k}}(a_{\mathbf{k}}-a_{\mathbf{k}}^{\dagger})
\end{equation}
or
\begin{equation} \label{hamilt}
H_{\rm{EA}}'=\frac{\gamma}{2}\sum_{\mathbf{k}}(a_{\mathbf{k}}a_{-\mathbf{k}}+a_{\mathbf{k}}^{\dagger}a_{-\mathbf{k}}^{\dagger}).
\end{equation}
The anisotropic Hamiltonians (\ref{hamil}) and (\ref{hamilt}) are called in the literature  Dzyaloshinsky-Moriya (DM) and exchange anisotropy (EA) interactions, respectively. Note that although $\gamma$ and $\gamma'$ can be very small, these terms cannot be considered perturbatively in the BEC - scenario especially in the region $H_{\rm{ext}}\sim H_{c}$, $T\sim T_{c}$, so one has to diagonalize the  Hamiltonian as a whole.

The effect of small $U(1)$ - symmetry breaking  within mean - field approximation (MFA) has been studied by Dell'Amore \textit{et al.}
\ci{andreasaniz} and Sirker \textit{et al.} \ci{sirker}
The authors of Ref. \ci{andreasaniz}  operated on a semi- classical level
and estimated the gap due to the anisotropy.


Sirker \textit{et al.} \ci{sirker} investigated the field-induced magnetic ordering transitions in TlCuCl$_{3}$ taking into account $H_{\rm{DM}}'$ as well as $H_{\rm{EA}}'$ within the framework of Hartree-Fock-Popov (HFP) approximation, which has been used to describe thermodynamic properties of quantum magnets in terms of  BEC - like physics with $U(1)$ symmetry.
Making an attempt to describe experimental magnetization curves $M(T,H)$
within HFP aproximation  they came to the following conclusions:
\begin{enumerate}
\item
The exchange anisotropy \re{hamilt} yields a small shift in condensed fraction but fails
to accurately describe experimental data;
\item
The DM anisotropy \re{hamil} has a dramatic effect even for $\gamma'\sim 10^{-3}\rm{meV}$
and smears out the phase transition into a crossover, i.e. there is no critical temperature above which the condensed fraction vanishes. However, it can explain only the experimantal  data on $M(T,H)$ for $H\|b$, but fails to accurately reproduce the data on $H\perp$ (1,0,\={2});
\item
The problem of an unphysical  jump in theoretical magnetization  curves may be solved by taking into account
DM anisotropy term and renormalization of the coupling constant.
\end{enumerate}

Thus     a complete theoretical description of experimental magnetization data of TlCuCl$_3$, together with the  phase diagram, i.e. $T_{c}(H)$,  is still missing \ci{zapf}, and
   a more sophisticated analysis beyond the HFP approximation is required for
 a better agreement with  the experimental data.
 In the present work, we propose an alternative MFA approach which gives a better  description of the magnetization data
 on   TlCuCl$_3$ including the  exchange anisotropy by using only three fitting parameters.

 To begin with, we have recently shown \ci{ourphysrev,ourannals}, in agreement with Refs. \ci{nikinu} and \ci{sirker} that the jump in the calculated  magnetization data at $T_c$
  is an artefact of the  HFP approximation, whereas the application of a more accurate approximation, e.g.  Hartee-Fock-Bogolyubov (HFB), can solve this problem.

 Another artifact of the HFP approximation is that it predits a discontinuty in the heat capacity, which was also noted by Dodds \textit{et al.} \ci{dodds} who applied this approximation to Ba$_3$Cr$_2$O$_8$,
 where $U(1)$ symmetry breaking is negligible.

In the present work we shall develop the  HFB approximation taking into account the exchange anisotropy term  $H_{\rm{EA}}'$. It is well known that the main difference between HFP and HFB approximation lies in consideration of the anomalous density-$\sigma$, which is completely neglected in the HFP but taken into account in the  HFB approximation. In our construction we assume that our formalism must coincide with  that of Sirker \textit{et al.} \ci{sirker} in the  particular case when $\sigma$ is set to zero.
We will show that in the system with a weakly explicitly broken U(1) symmetry
 the anomalous density $\sigma$
  may survive even at $T>T_{c}$ in contrast to the case with the  SSB.

The usage of the  HFB approximation even for the system with $U(1)$ symmetry
   has its own problem,
   which is called in the literature  the Hohenberg-Martin dilemma \ci{hohenberg}.
 Its content is the following: the theory, based on the standard grand canonical ensemble with SSB is internally inconsistent. Depending on the way of calculations, one obtains either a physical gap in the spectrum of collective excitations, or local conservation laws,
  together with general thermodynamic relations, become invalid. Recall that the excitation spectrum, according to the    Hugenholtz–-Pines theorem must be gapless \ci{pines} whereas the average of quantum fluctuation should be zero: $\langle a_{\mathbf{k}}\rangle=\langle a_{\mathbf{k}}^{\dagger}\rangle=0$. The solution of this dilemma was proposed by Yukalov and Kleinert  \ci{yukkl}, who suggested
  to introduce
   additional Lagrange multipliers
    \footnote{
    A similar version of MFA has been developed for
    disordered Bose systems and successfully applied to study
    the properties of Tl$_{1-x}$K$_x$CuCl$_3$ quantum magnets \ci{ournjp}.
    }.
      Assuming that our theory must  coincide in general with the HFB approximation of Ref. \ci{yukkl}, when $\gamma\rightarrow 0$, we shall  extend this method to the case of a weak anisotropy.

This paper is organized as follows. In Section II we revise the  Hohenberg - Martin dilemma which reveals the ambiguity of the  determination of the chemical potential in the SSB phase.
In Sect. III we will show that this ambiguity remains to  exist in the explicitly
U(1) symmetry broken phase and show how it may be overcome. In Sect. IV we apply our method
to TlCuCl$_3$ and show that it gives a good theoretical description of magnetization curves.  The Sect.V summarizes our results.

Below we adopt the units $k_{B}\equiv1$ for the Boltzmann constant, $\hbar\equiv1$
for the Planck constant, and $V\equiv1$ for the unit cell volume.
In these units the energies are
measured in Kelvin, the mass $m$ is expressed in \rm{K$^{-1}$}, the magnetic susceptibility
$\chi$ for the magnetic fields measured in Tesla has the units of \rm{K/T$^2$}, while the momentum
and specific heat $C_v$ are dimensionless. Particularly, the Bohr magneton is
$\mu_{B}={\hbar e}/{2m_{0}c}=0.671668$ \rm{K/T}, where $m_{0}$ is the free electron mass, and $e$ is the fundamental charge.


\section{ Hohenberg-Martin dilemma}
We start with the Hamiltonian
\begin{equation}
\fl
H=\int d^3r[\psi^{\dag}(\mathbf{r})(\hat{K}-\mu)\psi(\mathbf{r})+
\frac{U}{2}\left(\psi^{\dag}(\mathbf{r})\psi(\mathbf{r})\right)^{2}+\frac{{\gamma}}{2}\left(\psi^{\dag}(\mathbf{r})\psi^{\dag}(\mathbf{r})+\psi(\mathbf{r})\psi(\mathbf{r})\right)]
\lab{htot}
\end{equation}
where $\psi(\mathbf{r})$ is the Bosonic field operator, $U$ is the interaction strength and $\hat{K}$ is the kinetic energy operator which defines the bare triplon dispersion $\varepsilon_{k}$ in momentum space. The integration is performed over the unit cell of the crystal with the corresponding momenta defined in the first Brillouin zone. The parameter $\mu$ characterizes an additional direct contribution to the triplon energy due to the external field $H_{\rm{ext}}$,
\bea
\mu=g\mu_{B}H_{\rm{ext}}-\Delta_{\rm{st}}
\lab{muh}
\eea
and can be interpreted as a chemical potential  of the $S_{z}=-1$ triplons. In Eqs. \re{much1} and
 \re{muh} $g$   is the electron Land\'{e} factor and $\Delta_{st}$ is the spin gap separating the singlet ground state from the lowest-energy triplet excitations, $\Delta_{st}=g\mu_{B} H_{c}$, where $H_{c}$ is the critical field when the triplons start to form.

We assume that the exchange anisotropy is described by the last term in \re{htot}
 where the parameter ${\gamma}$ characterizes its strength.
It is clear that this term violates $U(1)$ symmetry, $\psi(\mathbf{r})\rightarrow e^{i\varphi}\psi(\mathbf{r})$ explicitly,
so strictly speaking there would be neither a Goldstone mode nor a Bose condensation \ci{yukan}. Nevertheless assuming $\tilde{\gamma}\equiv\gamma/U\ll1 $ is very small, one may make a
Bogolyubov shift in the field operator as
\be
\psi(\bf r)=\phi_{0}(\bf r)+\widetilde{\psi}(\bf r),
\lab{z}
\ee
%
%
where for the uniform case $\phi_0(\mathbf{r})$ is a real number.
 Note that when ${\gamma}=0$ and the U(1) symmetry is spontaneously broken,  $\phi_{0}(\mathbf{r})$ and $\widetilde{\psi}(\mathbf{r})$ are related to the density of condensed and uncondensed particles respectively. Following such an  interpretation we assume the orthogonality of the functions $\phi_{0}(\mathbf{r})$ and $\tilde{\psi}(\mathbf{r})$, i.e
\bea
\int d^3r\tilde{\psi}(\bfr)\phi_{0}(\bfr)=0
\eea
and for the simplicity call  the constant $\rho_{0}=\phi_{0}^{2}$  the density of condensed particles  \ci{yukan}. Similarly the quantity  $\rho_{1}=({1}/{V})\int d^3r\langle\tilde{\psi}^{\dag}(\mathbf{r})\tilde{\psi}(\mathbf{r})\rangle$  ,
  will be addressed as the density of uncondensed particles, so that the total number of particles
\bea
N=\int d \bfr\langle\psi^{\dag}(\bfr)\psi(\bfr)\rangle
\eea
defines the density of triplons per unit cell $\rho=N/V=\rho_{0}+\rho_{1}$.

The total magnetization per site is associated with  the number of triplons as    $M=g\mu_{B}N $
and the transverse one is  $M_{\perp}=g\mu_{B}\sqrt{\rho_0/2}$ \ci{zapf}.
Below we assume that there is a critical temperature defined as $\rho_{0}(T_{c})=0$, so that , $\rho_{0}(T\geq T_{c})=0$ and $\rho(T\geq T_{c})=\rho_{1}$. Clearly due to the anisotropy the energy spectrum has a gap in both phases.

Now we apply the standard technique used in the HFB formalism \ci{ourannals} and start with the Fourier transformation for quantum fluctuations
\bea \label{q}
\tilde{\psi}(\mathbf{r})=\sum_{\mathbf{k}}a_{\mathbf{k}}e^{i\mathbf{k}\mathbf{r}}.
\eea
The summation by momentum, which should not include $\bfk=0$ states, may be replaced
by momentum integration as it is outlined in the  Appendix A.

After using $(\ref{z})$ and $(\ref{q})$ the Hamiltonian \re{htot} is presented as the sum of five terms
\bea \label{h}
H=\sum_{n=0}^{4}H_{n} ,
\eea
labeled according to their order with respect to $a_{\mathbf{k}}$ and $a_{\mathbf{k}}^{\dag}$.
The zero-order term does not contain field operators of uncondensed triplons
\bea
H_{0}=-\mu\phi_{0}^{2}+{\gamma}\phi_{0}^{2}+\frac{U}{2}\phi_{0}^{4}.
\eea
The linear term is
\bea
H_{1}=\sum_{\mathbf{k}}\{a_{\mathbf{k}}^{\dag}\sqrt{\rho_{0}}({\gamma}
-\mu+U\rho_{0})+\rm{h.c.}\}\delta_{\mathbf{k},0},
\eea
the quadratic term is
\bea
H_{2}=\sum_{\mathbf{k}}(\varepsilon_{k}-\mu+2U\rho_{0})a_{\mathbf{k}}^{\dag}a_{\mathbf{k}}
+\frac{U}{2}(\widetilde{\gamma}+\rho_{0})\sum_{\mathbf{k}}(a_{\mathbf{k}}a_{-\mathbf{k}}+a_{\mathbf{k}}^{\dag}a_{-\mathbf{k}}^{\dag}),
\eea
where $\widetilde \gamma=\gamma/U$,
and the third and forth order terms are given by
\be
\ba
H_{3}=U\sqrt{\rho_{0}}\sum_{\mathbf{k},\mathbf{p}}[a_{\mathbf{p}}^{\dag}a_{\mathbf{p}-\mathbf{k}}a_{\mathbf{k}}+ a_{\mathbf{k}}^{\dag}a_{\mathbf{p}-\mathbf{k}}^{\dag}a_{\mathbf{p}}],\\
H_{4}=\displaystyle{\frac{U}{2}}\sum_{\mathbf{k},\mathbf{p},\mathbf{q}}a_{\mathbf{k}}^{\dag}a_{\mathbf{p}}^{\dag}a_{\mathbf{q}}a_{\mathbf{k}+\mathbf{p}-\mathbf{q}} \quad.
\ea
\ee
To diagonalize $H$ we use following prescription, based on the Wick theorem:
\be
\ba
\fl
 a_{\mathbf{k}}^{\dag}a_{\mathbf{p}}a_{\mathbf{q}}\rightarrow 2\langle a_{\mathbf{k}}^{\dag}a_{\mathbf{p}}\rangle a_{\mathbf{q}}+a_{\mathbf{k}}^{\dag} \langle a_{\mathbf{p}}a_{\mathbf{q}}\rangle, \\
\fl
a_{\mathbf{k}}^{\dag}a_{\mathbf{p}}^{\dag}a_{\mathbf{q}}a_{\mathbf{m}}\rightarrow4a_{\mathbf{k}}^{\dag}a_{\mathbf{m}}\langle a_{\mathbf{p}}^{\dag}a_{\mathbf{q}}\rangle
+a_{\mathbf{q}}a_{\mathbf{m}}\langle a_{\mathbf{k}}^{\dag}a_{\mathbf{p}}^{\dag}\rangle+a_{\mathbf{k}}^{\dag}a_{\mathbf{p}}^{\dag}\langle a_{\mathbf{q}}a_{\mathbf{m}}\rangle
-2\rho_{1}^{2}-\sigma^{2},
\lab{wick}
\ea
\ee
where $\langle a_{\mathbf{k}}^{\dag}a_{\mathbf{p}}\rangle=\delta_{\mathbf{k},\mathbf{p}}n_{\mathbf{k}}$, \quad $\langle a_{\mathbf{k}}a_{\mathbf{p}}\rangle=\delta_{\mathbf{k},-\mathbf{p}}\sigma_{\mathbf{k}}$   with $n_{\mathbf{k}}$ and $\sigma_{\mathbf{k}}$ being related to the normal $(\rho_{1})$, and
 anomalous $(\sigma)$ densities as
\be
\rho_{1}=\sum_{\mathbf{k}}n_{\mathbf{k}}=\sum_\mathbf{k} \langle a_{\mathbf{k}}^{\dag}a_{\mathbf{k}}\rangle ,
\lab{rho11}
\ee
\be
     \sigma=\sum_{\mathbf{k}}\sigma_{\mathbf{k}}=\frac{1}{2}\sum_\mathbf{k}
     \left(\langle a_{\mathbf{k}}a_{-\mathbf{k}}\rangle+\langle a_{\mathbf{k}}^{\dag}a_{-\mathbf{k}}^{\dag}\rangle\right) .
     \lab{sigma1}
\ee
Here we underline that the main difference between the HFP and HFB approximations concerns
the anomalous density: neglecting
$\sigma$ as well as $\langle a_{\bfk} a_{\bf p}\rangle$ in \re{wick} one
arrives at the HFP approximation,
which can also be obtained in variational perturbation theory \ci{Kleinert05}.
However, the normal, $\rho_{1}$, and anomalous averages, $\sigma$, are equally important and
neither of them can be neglected without making the theory not self-consistent \ci{AndersenRevModPhys,yukanom,Yukalov11}.
Although $\rho_1$ and $\sigma$ are functions of temperature and external magnetic field, we omit their explicit dependence in the formulas to avoid  confusion.

This approximation simplifies the Hamiltonian $(\ref{h})$ as follows
\bea \label{ham}
H=H_{0}+H_{\rm{lin}}+H_{\rm{bilin}} ,
\eea
\bea \label{hc}
H_{0}=-\mu\rho+{\gamma}\rho_{0}+\frac{U}{2}
\rho_{0}^{2}-\frac{U}{2}(2\rho_{1}^{2}+\sigma^{2})  ,
\eea
\be
H_{\rm{lin}}=\sqrt{\rho_{0}}\sum_{\mathbf{k}}\{a_{\mathbf{k}}^{\dag}[
{\gamma}-\mu+\rho_{0}U+2\rho_{1}U+\sigma U]+\rm{h.c.}\}  ,
\lab{hlin}
\ee
\bea
H_{\rm{bilin}}=\sum_{\mathbf{k}}(\varepsilon_{k}-\mu+2U\rho)a_{\mathbf{k}}^{\dag}a_{\mathbf{k}}+
\frac{U(\widetilde{\gamma}+\rho_{0}+\sigma)}
{2}\sum_{\mathbf{k}}(a_{\mathbf{k}}a_{-\mathbf{k}}+a_{\mathbf{k}}^{\dag}a_{-\mathbf{k}}^{\dag}).
\eea
From ($\ref{hlin}$), requiring $H_{\rm{lin}}=0$ \ci{stoofbook}  we  obtain the following equation for $\mu$
\bea \label{myu1}
\mu=U[\rho_{0}+2\rho_{1}+\sigma+\widetilde{\gamma}] .
\eea
It can be shown \ci{ourlat} that  the minimization of the
thermodynamic potential $\Omega$ with respect to $\rho_0$, i.e. using the equation  $\partial\Omega/\partial\rho_{0}=0$  leads to the same equation as ($\ref{myu1}$).

The Hamiltonian ($\ref{ham}$) can be easily diagonalized by implementing a Bogolyubov transformation. We  refer the reader to the Appendix B for details and present here only the main results  of this procedure, valid both for $T\leq T_{c}$ and $T>T_{c}$ cases.

a) The quasiparticle  dispersion
\bea \label{ek}
E_{k}=\sqrt{(\varepsilon_{k}+X_{1})(\varepsilon_{k}+X_{2})} .
\eea
b) Main equations
\bea \label{x12}
\left\{
  \begin{array}{ll}
    X_{1}=-\mu+U[2\rho+\widetilde{\gamma}+\rho_{0}+\sigma] & \hbox{} \\
    X_{2}=-\mu+U[2\rho-\widetilde{\gamma}-\rho_{0}-\sigma] & \hbox{} .
  \end{array}
\right.
\eea
c) Normal and anomalous self energies
\bea \label{sig}
\fl
\Sigma_{n}=2U\rho=(X_{1}+X_{2})/2+\mu, \quad  \quad \Sigma_{an}=U[\sigma+\rho_{0}+\widetilde{\gamma}]=(X_{1}-X_{2})/2 .
\eea
d) Normal and anomalous densities
\bea \label{ro}
\rho_{1}=\frac{1}{V}\displaystyle{\sum_{\mathbf{k}}}\left\{\frac{W_{k}
[\varepsilon_{k}+(X_{1}+X_{2})/2]}{E_{k}}-\frac{1}{2}\right\} ,
\eea
\bea \label{sigm}
\sigma=\frac{X_{2}-X_{1}}{2V}\sum_{\mathbf{k}}\frac{W_{k}}{E_{k}} ,
\eea
where $W_{k}=\coth(E_k/2)/2=f_{B}(E_{k})+{1}/{2}$,
$f_{B}(x)=1/(e^{\beta x}-1)$ .

Now we are ready to illustrate the Honenberg-Marting dilemma which
applies to  the spontaneous symmetry broken (SSB) phase, when ${\gamma}$=0 and $T\leq T_{c}$.

{\bf SSB case.} In this phase we have the  Hugenholtz-Pines  theorem \ci{pines}:
\bea
\Sigma_{n}-\Sigma_{an}=\mu .
\eea
From equations ($\ref{sig}$) one obtains
\bea
\Sigma_{n}-\Sigma_{an}=X_{2}+\mu .
\eea
 Clearly, this theorem is satisfied when $X_{2}=0$. Note that this condition leads automatically to the gapless energy dispersion:
\bea
E_{k}\mid_{X_{2}=0}=\sqrt{(\varepsilon_{k}+X_{1})\varepsilon_{k}}=ck+O(k^{3}) .
\eea
On the other hand we may set in ($\ref{x12}$) $X_{2}=0$ and $\gamma=0$ to obtain
\bea \label{myu2}
\mu=U[2\rho-\rho_{0}-\sigma]=U[2\rho_{1}+\rho_{0}-\sigma] .
\eea
Comparing both chemical potentials given in ($\ref{myu1}$), for $\tilde\gamma=0$, and ($\ref{myu2}$) with each other one may make the  following conclusions:
\begin{itemize}
 \item
 The chemical potentials  are the same in HFP approximation when $\sigma=0$. So,  there is no Hohenberg - Martin dilemma in this approximation and hence,  the usage of the requirement $H_{\rm{lin}}=0$
 by Sirker \textit{et al.} \ci{sirker} is justified.
\item
However, when $\sigma$ is taken into account
i.e. when one is dealing with the  HFB approximation , they are different.
\end{itemize}
In other words, in the SSB phase, the conditions $H_{\rm{lin}}=0$ and
$\Sigma_{n}-\Sigma_{an}=\mu$ are consistent in the  HFP but not in the HFB approximations.
This contradiction  is the essence of the Hohenberg  - Martin dilemma.
So, when $\sigma$ is taken into account
 one can choose only one of the two requirements  $H_{\rm{lin}}=0$ or $\Sigma_{n}-\Sigma_{an}=\mu$ in the SSB phase.
The solution of this problem has been proposed by Yukalov and Kleinert \ci{yukkl} recently.

\section{HFB approximation for explicitly broken $U(1)$ phase}

Following Ref. \ci{yukkl} we introduce   two Lagrange multipliers, say $\mu_{0}$ and $\mu_{1}$. One of them guarantees the requirement $H_{\rm{lin}}=0$,
 or equalently $d\Omega/d\rho_0=0$,
 while the second one is chosen in order to satisfy Hugenholtz-Pines theorem.
Using ($\ref{myu1}$) and ($\ref{myu2}$) we define
\bea
\mu_{0}=U[2\rho_{1}+\rho_{0}+\sigma] ,
\eea
\bea
\mu_{1}=U[2\rho_{1}+\rho_{0}-\sigma] ,
\eea
for the SSB  case.
The whole physical chemical potential, $\mu$, which is related to the free energy as $N=-({\partial\Omega}/{\partial\mu})_V$ is given by
\bea \label{mufull}
\mu=(\mu_{0}\rho_{0}+\mu_{1}\rho_{1})/\rho ,
\eea
so that  $N_{0}=-\left({\partial\Omega}/{\partial\mu_{0}}\right)_V$ and $N_{1}=-\left({\partial\Omega}/{\partial\mu_{1}}\right)_V$. Clearly, in  the normal phase $\rho_0=0$ and , hence, $\mu=\mu_1$.

Now we may come back to develop a theory for a more  general case with finite  exchange anisotropy, assuming that it must coincide with the Yukalov-Kleinert HFB approximation in the particular case when ${\gamma}=0$. In other words the SSB case  will be our benchmark.

Following the  Yukalov-Kleinert prescription one may rewrite equations ($\ref{myu1}$) and ($\ref{x12}$) as follows:
\footnote{See Appendix B.}
\bea \label{mueta1}
\mu_0=U[2\rho_{1}+\rho_{0}+\sigma+\widetilde{\gamma})] ,
\eea
\bea \label{x1eta1}
X_{1}=-\mu_{1}+U[2\rho+\widetilde{\gamma}+\rho_{0}+\sigma] ,
\eea
\bea \label{x2eta2}
X_{2}=-\mu_{1}+U[2\rho-\widetilde{\gamma}-\rho_{0}-\sigma] .
\eea
The equations \re{ek}, \re{ro} and  \re{sigm} remain formally unchanged.

\subsection{Condensed phase $T\leq T_{c}$.}

Bearing in mind that $\rho=\rho_{0}+\rho_{1}$, $\rho_{1}=\rho_{1}(X_{1},X_{2})$, $\sigma=\sigma(X_{1},X_{2})$  given by ($\ref{ro}$) and ($\ref{sigm}$), one notes that the system of equations  ($\ref{mueta1}$)-($\ref{x2eta2}$) is underdetermined.
In fact, with a given $\mu$ in ($\ref{mufull}$) we have three equations with respect to
 four unknown quantities: $X_{1}$, $X_{2}$, $\mu_{0}$ and $\rho_{0}$. In the ordinary HFB
 approximation with $\gamma=0$ this problem is solved by means of
 the  Hugenholtz-Pines theorem namely by  setting $X_{2}=0$ by hand and introducing an
   additional equation    as
\bea \label{sigma}
\Sigma_{n}-\Sigma_{an}=\mu_{1} .
\eea
However, when the anisotropy is included, the Hugenholtz -Pines
 theorem  no longer holds, and hence we have no right to use ($\ref{sigma}$) directly. On the other hand  ${\gamma}$ is assumed to be a small parameter of the system. So we naturally assume that, when the U(1) symmetry is slightly broken explicitly, the Hugenholtz -Pines theorem may be violated  up to terms  linear in ${\gamma}$.
Thus, by taking into account only a linear correction in $\gamma$ we assume
\bea \label{sigmac}
\Sigma_{n}-\Sigma_{an}=\mu_{1}+2{\gamma}c_{\gamma} ,
\eea
that is in the SSB phase, when  ${\gamma}=0$, the theorem will still  be exact.

In ($\ref{sigmac}$) $c_{\gamma}$ is a coefficent in the expansion of
$\Sigma_{n}-\Sigma_{an}$ in powers of $\gamma$ which can be fixed  e.g.
by  fitting the gap in the energy spectrum observed experimentally  at small momentum transfer. In other words we propose an additional equation ($\ref{sigmac}$) to have the complete  system of four equations   \re{sig},\re{ro}, ($\ref{x1eta1}$), and ($\ref{x2eta2}$)   with respect to four quantities: $X_{1}$, $X_{2}$, $\rho_{1}$ and $\rho_{0}$.
Now inverting \re{x2eta2} and using  (\ref{sig})  where $\mu$ is replaced
by $\mu_1$ we obtain
\bea \label{x2}
X_{2}=\Sigma_{n}-\Sigma_{an}-\mu_{1}=2{\gamma}c_{\gamma} .
\eea
Inserting this in ($\ref{x2eta2}$) gives
\bea \label{mux2}
\mu_{1}=U[\rho_0+2\rho_{1}-\sigma-\widetilde{\gamma}(1+2c_{\gamma})] ,
\eea
where we omitted higher terms of the  order $O(\gamma^2)$.
From ($\ref{mux2}$) and ($\ref{x1eta1}$) one defines $X_{1}$ as
\bea \label{x1rho}
X_{1}=2U[\rho_0+\sigma+\widetilde{\gamma}(1+c_{\gamma})],
\eea
( we remind here  that $\widetilde \gamma=\gamma/U$).

The excitation energy has a gap due to ${\gamma}$
\bea \label{gap}
E_{k}=\sqrt{(\varepsilon_{k}+X_{1})(\varepsilon_{k}+2{\gamma}c_{\gamma})}, \quad  E_{k}|_{k\rightarrow 0}=\sqrt{2X_{1}\gamma c_{\gamma}} .
\eea
To make a comparision with the HFP aproximation with anisotropy, as developed
by Sirker \textit{et al.} \ci{sirker},  we note that, in their approximation, the requirement
$H_{\rm{lin}}=0$ directly leads to
$X_2=2\gamma$  and  $\Sigma_{11}-\Sigma_{12}=\mu+2\gamma$, which is consistent with present  approach.

In contrast to cold  atomic gases,  the total number of particles
in the present triplon problem is an unknown quantity while the chemical potential serves as an  input parameter. So, excluding $\rho_0$ from Eqs.  (\ref{mux2}) and (\ref{x1rho})   we have
\be
\ba
\Delta_1\equiv\dsfrac{X_1}{2}=\mu_1+2U(\sigma-\rho_1)+\gamma(2+3c_{\gamma}) ,
\label{delta}
\ea
\ee
or introducing the  dimensionless variable $Z=\Delta_{1}/\mu_1$ and using
\re{sig},\re{ro} we obtain
\bea \label{zet1}
Z=1+\widetilde\sigma-\widetilde\rho_{1}+\frac{\gamma}{2U\rho_{c}^{0}}(2+3c_{\gamma}) ,
\eea
\bea \label{rho1}
\widetilde\rho_{1}
\equiv\widetilde\rho_{1}(Z)=\frac{\rho_1 (Z)} {\rho_{c}^{0}}=
\frac{1}{\rho_{c}^{0}}
\sum_{\mathbf{k}}\left\{\frac{W_{k}({\varepsilon}_{k}+\Delta_1+
\Delta_2)}
{{E}_k}-\frac{1}{2}\right\} ,
\eea
\bea \label{sigmaz}
\widetilde\sigma\equiv\widetilde\sigma(Z)=\frac{\sigma (Z)} {\rho_{c}^{0}}=
\frac{\Delta_2-\Delta_1}{\rho_{c}^{0}}\sum_\mathbf{k}\frac{W_k}
{{E}_k} ,
\eea
where
$\rho_{c}^{0}=\mu_1/2U$, $\mu_1=g\mu_B(H_{\rm{ext}}-H_c)$,
$E_{k}=\sqrt{({\varepsilon}_{k}+2\Delta_1)({\varepsilon}_{k}+2\Delta_2)}$,
and $\Delta_2=\gamma c_{\gamma} $.

The strategy of the numerical  calculations in the  $T\leq T_c$ phase is as follows:
Starting
with input parameters $T$, $H_{\rm{ext}}$,  $U$, ${\gamma}$ and $c_{\gamma}$, as well as the  parameters of the  bare dispersion (\ref{aq}) one solves the nonlinear algebraic equation (\ref{zet1}), where $\rho(Z)$ and $\sigma(Z)$ are given by (\ref{rho1}), (\ref{sigmaz}), with respect to $Z$, and then by using this solution, say, $Z_0$ in the following equation
 \be
 \rho_0=2Z_0\rho_{c}^{0}-\sigma(Z_0)-\widetilde\gamma(1+c_{\gamma})
 \label{rho0}
 \ee
    determines the  density of condensed particles.  The number of total particles may be found from $\rho=\rho_0+\rho_1(Z_0)$ where $\rho_1(Z_0)$ is evaluated by Eq. \re{rho1}.

\subsection{The critical temperature and triplon density .}

Clearly the total number of triplons $N=\rho V$ and among them the number of condensed ones $N_0=\rho_0 V$ depend on the external magnetic field, $H_{\rm{ext}}$ and the temperature $T$. For a given
$H_{\rm{ext}}>H_c$ there may be a critical point, $T=T_c$ where the condensed particles vanish. Lets formally define this temperature as a  critical temperature $T_c$, where $\rho_0(T\geq T_c)=0$ and
the value of the density at this point, $\rho(T_c)=\rho_{c}$ as a critical density.   To determine these quantities we use the approximation developed in the previous section.

Thus near $T_c$ the Eq. \re{mux2} can be rewritten as
\bea
\mu_1(T\rightarrow T_c)=U[2\rho_c-\sigma_c-\widetilde{\gamma}(1+2c_{\gamma})]=
g\mu_BH_{\rm{ext}}-\Delta_{\rm{st}}.
\eea
and hence,
\be
\ba
\rho_{c}=\dsfrac{g\mu_BH_{\rm{ext}}-\Delta_{\rm{st}}}{2U}+
\dsfrac{\sigma_c+\widetilde{\gamma}(1+2c_{\gamma})}{2}\equiv
\rho_{c}^{0}+\frac{\sigma_c+\widetilde{\gamma}(1+2c_{\gamma})}{2} .
\lab{rhoc2}
\ea
\ee
The energy dispersion of phonons becomes
\bea \label{Ekc}
E^c_k=E_k(T\rightarrow T_c)=\sqrt{(\varepsilon_k+X^c_1)
(\varepsilon_k+2{\gamma}c_{\gamma})} ,
\eea
where $X^c_1$, by using  \re{x1rho}, is given as
\bea \label{x1c}
X^c_1=X_1\mid_{T\rightarrow T_c}=2U[\sigma_c+\widetilde{\gamma}(1+c_{\gamma})] .
\eea
Inserting these expressions into   \re{ro},  \re{sigm} and using \re{rhoc2} one finds the critical temperature by solving the  following nonlinear algebraic  equations
with respect to $T_c$ and $\sigma_c$
\numparts \begin{eqnarray}\label{eqscrit}
\fl
\sum_{\mathbf{k}}\dsfrac{f_B(E^c_k)}{E^c_k}[\varepsilon_k+
U(\sigma_c+\widetilde{\gamma}(1+2c_{\gamma}))]=
\dsfrac{g\mu_BH_{\rm{ext}}-\Delta_{\rm{st}}}{2U}+
\dsfrac{\sigma_c+\widetilde{\gamma}(1+2c_{\gamma})}{2} ,  \label{eqscrita}\\
\sigma_c=-U(\sigma_c+\widetilde{\gamma})\ds\sum_\mathbf{k}\dsfrac{f_B(E^c_k)}{E^c_k} ,
   \label{eqscritb}
\end{eqnarray} \endnumparts
 where $f_B(E^c_k)=1/(\exp(E^c_k/T_c)-1)$. Having solved these equations the critical density may be evaluated by using \re{rhoc2}, where $\rho_{c}^{0}$ is the critical density
 at $\gamma=0$, i.e. $\rho_{c}^{0}=\rho_c(\gamma=0)=\mu_1/2U$.

 A natural question here arises: Does the anomalous density $\sigma$ survive at $T\geq T_c$? To answer this question we first consider the limiting simpler case with ${\gamma}=0$

{\bf {SSB  phase: $\gamma=0$} .}

When ${\gamma}=0$,   Eq. \re{eqscritb} becomes
\bea
\label{siggc}
\sigma_c=-U\sigma_c\sum_\mathbf{k}\frac{f_B(E^c_k)}{E^c_k}\equiv-\sigma_c {A_c} ,
\eea
where
\bea \label{sigmgam=0}
{A_c}=U\sum_\mathbf{k}\frac{f_B(E^c_k)}{E^c_k} .
\eea
Since $U>0$ the only solution of (\ref{siggc}) is  trivial:
\be
\sigma_c\mid_{{\gamma}=0}=0
\ee
and hence from \re{rhoc2}  and \re{eqscrita} one may obtain the  familiar equation \ci{ourannals}:
\bea \label{rhoc0}
\rho^{0}_{c}=\frac{\mu}{2U}=\sum_\mathbf{k}\frac{1}{e^{\varepsilon_k/T^{0}_{c}}-1}
=\frac{g\mu_B H_{\rm{ext}}-\Delta_{\rm{st}}}{2U} ,
\eea
to calculate the critical temperature $T^{0}_{c}$ of the system in the  isotropic case.

{{\bf Explicitly broken symmetry phase: ${\gamma}\neq$0}.}

Now Eq. \re{eqscrit} has a formal solution for $\sigma_c$
\bea
\sigma_c=-\frac{{\widetilde\gamma}A_c}{1+A_c} ,
\eea
where $A_c$ is given in (\ref{sigmgam=0}).
Since $A_c$ is finite, it is seen that for a system with exchange anisotropy ${\gamma}\neq 0$, the anomalous density is also finite even at $T=T_c$ in contrast to the SSB case.
Note that, in general,  $\sigma$  is negative as stated by Griffin \ci{griffin}.
 For numerical evaluations it is convenient
to search $\sigma_c$ and $\rho_c$ as  $\sigma_c=-\gamma\sigma_x/U$ and
$\rho_c=\rho_x\rho_{c}^{0}$ where now $\rho_x$ and $\sigma_x$ will be real  numbers
the  order of $0.5 ..2$.
Therefore one may conclude that $\sigma_c=0 $ if $\gamma=0$ and $\sigma_c\neq 0 $ otherwise.
Actually, as we will show in the next  section  $\mid\sigma_c\mid$ is rather small.

The Eqs.  \re{eqscrit} and \re{rhoc0}     may be used to  estimate  the shift of critical temperature due to the exchange anisotropy
\bea
\frac{\Delta T_c}{T^0_c}\equiv\frac{T_c-T^0_c}{T^0_c}\quad,
\eea
where $T^0_c=T_c(\gamma=0)$ is the critical temperature for the $\gamma=0$ case defined by  Eq. \re{rhoc0}.

\subsection{Normal phase $T>T_c$.}
In the normal phase, $\rho_0=0$, $\rho_1=\rho$, $\mu_1=\mu$ and $T>T_c$,
the energy dispersion has a gap even for $\gamma=0$ and the equations  (\ref{mux2}), (\ref{x1rho}) are no longer valid. However the main equations  (\ref{x12}) with $\rho_0=0$
\numparts \begin{eqnarray}
X_{1}=-\mu+U[2\rho+\sigma+\tilde{\gamma}] \label{xeta12a} ,\\
    X_{2}=-\mu+U[2\rho-\sigma-\tilde{\gamma}] \label{xeta12b} ,
\end{eqnarray} \endnumparts
 make sense, where $\mu$ is defined in \re{muh}.
The normal and anomalous self energies are
\bea \label{sig1}
\Sigma_{n}=(X_{1}+X_{2})/2+\mu=2U\rho
\eea
and
\bea \label{sig2}
 \Sigma_{an}=(X_{1}-X_{2})/2=U(\sigma+\widetilde{\gamma}) .
\eea
Clearly,  the Hugenholtz–-Pines theorem is not valid
\bea
\Sigma_{n}-\Sigma_{an}=2U\rho-U(\sigma+\widetilde{\gamma})\neq\mu
\eea
even for the  ${\gamma}=0$ case.
Similarly to the previous subsection it can be shown that $\sigma(T> T_c)\neq0$
being defined as
\bea
\sigma=-\frac{A\widetilde{\gamma}}{1+A} ,
\eea
where
\bea \label{sigmgamT}
A=U\sum_\mathbf{k}\frac{f_B(E_k)}{E_k} ,
\eea
with
\bea \label{EkcT}
E_{k}=\sqrt{(\varepsilon_{k}+X_{1})(\varepsilon_{k}+X_{2})}=\sqrt{(\varepsilon_k-\mu^0_{eff})^2-U^2(\sigma+\widetilde{\gamma})^2} .
\eea
The density of triplons is
\bea \label{roT}
\rho=\rho_{1}=\sum_{\mathbf{k}}\frac{(\varepsilon_k-\mu^0_{eff})}{E_{k}}f_B(E_k) ,
\eea
where we used (\ref{ro}) and introduced the effective chemical potential $\mu^0_{eff}=\mu-2U\rho$.

Thus, we have to solve the system of two algebraic nonlinear equations with respect to $\rho_x$ and $\sigma_x$
\numparts \begin{eqnarray}
\rho_x-\dsfrac{1}{\rho_{c}^{0} }\sum_\mathbf{k}\dsfrac{\veps_k}{E_k}
-2A(\rho_x-1)=0 ,\label{rhosigtba} \\
 \sigma_x-\dsfrac{A}{1+A}=0 , \label{rhosigtbb}
\end{eqnarray} \endnumparts
defined as $\rho=\rho_x\rho_{c}^{0}$ and $\sigma=-\sigma_x\widetilde\gamma$.
 In Eq.  \re{rhosigtba} $\rho^0_c=\mu/2U$,
 $E_k=\sqrt{[\veps_k-\mu(1-\rho_x)]^2-\gamma^2(1-\sigma_x)^2}$ with $\mu $ is given in Eq. \re{muh}.
One may see that the presence of anisotropy leads to a state with $\rho_0=0$ but $\sigma\neq0$, which is in contrast to the case with ${\gamma}=0$ in the HFB approximation, where $\sigma(T\geq T_c)=0$. We will discuss this point in next section.

\section{Results and discussions.}
Among all quantum magnets, the compound TlCuCl$_3$ is well known for
its rather large $U(1)$ symmetry breaking \ci{zapf}. Therefore it is a good example
to apply the present approach. Experimental data on magnetization curves $M(T,H)$ as well as on the phase boundary $T(H)$ for TlCuCl$_3$ have been reported
in Refs. \ci{oosawa,yamada,delamore} in the range of 5\rm{T}$\leq$$H$$_{\rm{ext}}$$\leq$8\rm{T}, 2K$\leq$$T$$<$7\rm{K}. As it was pointed out in the introduction,
 the previous theoretical description of these data, mostly based on the  HFP approximation, is good
  only for the $H\parallel b$ case when a DM anisotropy is included \ci{sirker}.
  Moreover, in general, this approximation leads to  an unphysical jump in the magnetization near the transition point. It has been shown that this artefact cannot be improved  neither by using a more realistic dispersion relation \ci{misguich} nor by taking into account an exchange anisotropy \ci{sirker}.  We have recently agrued that this artefact is a characteristic feature of the HFP approximation caused by neglecting the  anomalous density $\sigma$ \ci{ourannals}. However, in  Ref.\ci{ourannals} we  did not make an attempt to compare our results with the experiments since  anisotropy effects were not taken into account.
In the previous sections of this paper  we have developed a MFA where the anomalous density as well as
the exchange anisotropy term \re{hamilt} are included .
In this section we shall use  this approach for a  theoretical
description of the magnetic properties  of  TlCuCl$_3$.

First we argue that for  the bare dispersion $\veps_k$
 given in \ci{misguich} (see appendix A) is the most preferable one.
 In fact, by choosing quadratic or relativistic \ci{shermanprl} bare dispersions
 one usually performs integration by momentum in the whole space $(k_x,k_y,k_z=-\infty ... \infty)$ , while the choice of the realistic dispersion \ci{misguich} implies an
   integration in the first Brilloine zone.

 Having fixed the dispersion  relation  we are left with only three parameters
 $\gamma$, $c_{\gamma}$ and $U$. As for the g-factor, we may use available experimental data
  where the g - factor  is reported as $g=2.23$  for $H_{ext}\perp (1,0,$\={2}$)$
  and $g=2.06$ for $H_{ext}\parallel $ b. To optimize these free parameters we used the experimental  phase diagram and  magnetization curves  from
   Ref. \ci{yamada}. The result of the corresponding fits are $\gamma=0.045K$, $c_{\gamma}=1.67$ and $U=367.5K$.

 In Figs. 1 (a) and (b) we show a comparision between the experimental data and the resulting
 fits to these data.
\begin{figure}[H]
\begin{minipage}[H]{0.49\linewidth}
\center{\includegraphics[width=1.1\linewidth]{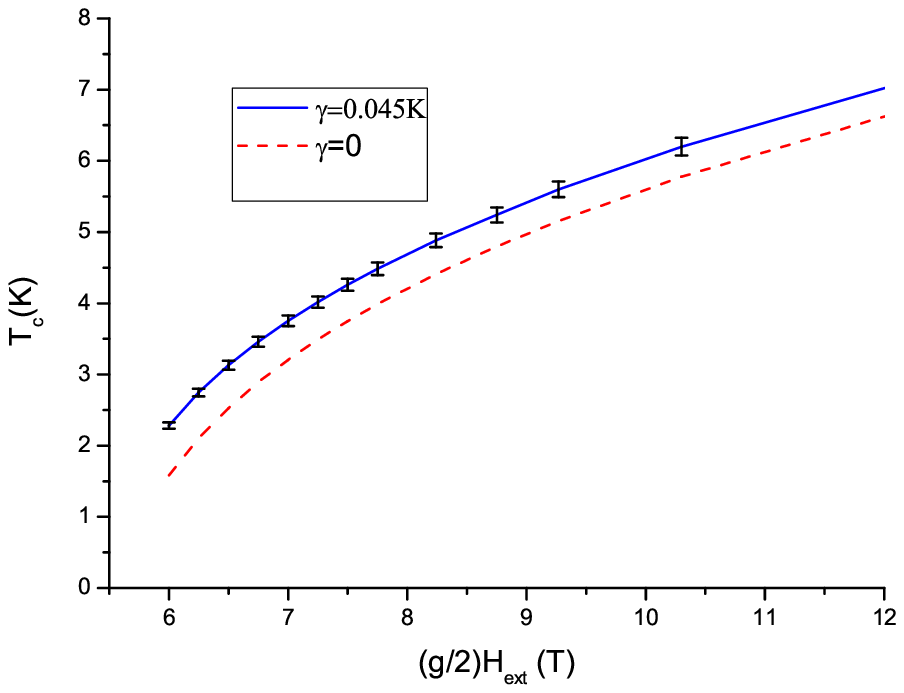} \\ a)}
\end{minipage}
\hfill
\begin{minipage}[H]{0.49\linewidth}
\center{\includegraphics[width=1.1\linewidth]{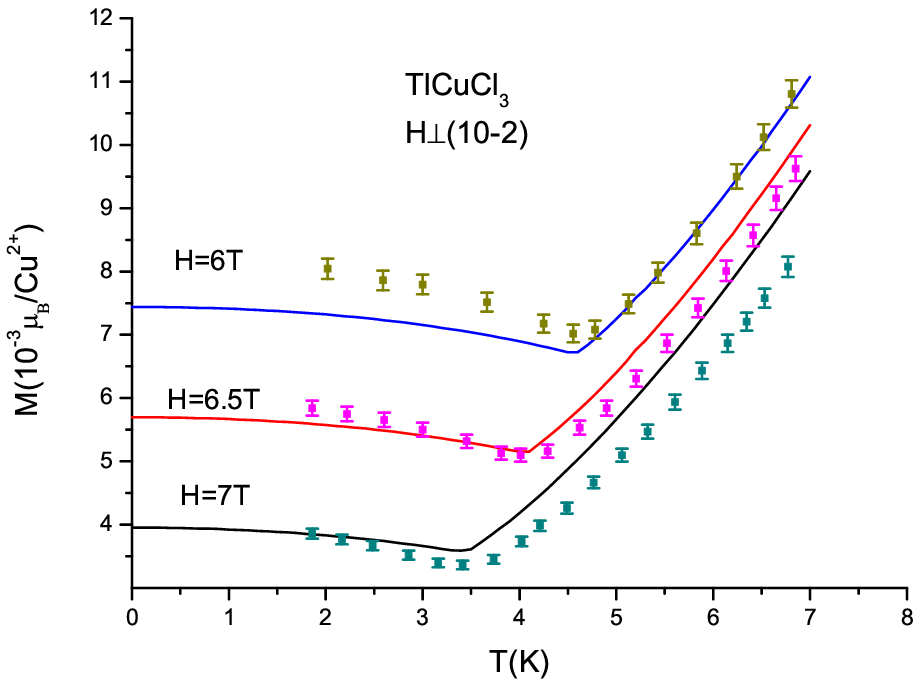} \\ b)}
\end{minipage}
\caption{
Phase diagram normalized by the g-factor (a) and low-  temperature magnetization curves
 (b) for TlCuCl$_{3}$ . The experimental data are taken from
 \ci{yamada}. The input parameters
 are   $\gamma=0.045\rm{K}$,  $c_{\gamma}$=1.67 , U=367.52\rm{K} , g=2.23 and $\Delta_{\rm{st}}=7.54\rm{K}$
 for our HFB approximation
}
\label{fig1ab}
\end{figure}

 The phase boundary  $T_c(H_{\rm{ext}})$, displayed in Fig.1 (a), is intersting by itself, since
 it contains information about the critical exponent $\phi$, defined as $T_c\sim(H_{\rm{ext}}-H_c)^\phi$ or more precisely as $T_c=\rm{const}\times(\mu/U)^\phi$, where
 the constant   and $\phi$ are fitting parameters.  Note that for the case of a  homogeneous
 ideal gas with the quadratic  dispersion $\phi=2/3$. From  Fig.1(a) we have found that
 $\rm{const}=47.4 \rm{K}$ and $\phi=0.53$  (solid line)  and $\rm{const}= 63.2 \rm{K}$ and $\phi=0.62$  (dashed line)
 for the $\gamma=0.045 \rm{K}$ and $\gamma=0$ cases respectively.
 This means that the inclusion of a finite
 exchange anisotropy reduces the exponent $\phi$, and one does not need
  to  expect $\phi = 2/3$ as it has been debated in the literature \ci{shermanprl,misguich,yamadaglass}. In fact, the presence of the interparticle interaction as well
 as using a more realistic dispersion than a simple quadratic one, leads
 to a shift of the critical temperature, especially at high temperatures $T>2\rm{K}$ \ci{kastening}. Here we note that, if we restrict to fit
  $\phi$ in the range of
 $0\leq T\leq 1.5\rm{K}$ (not drawn in Fig.1a ) the solid line in Fig.1 (a) may also
  be well fitted by $\phi\approx 2/3.$

 In Fig. 1(b) the magnetization curves for various $H_{\rm{ext}}$ are presented
   in
 comparison with the experimental data from Ref. \ci{yamada}  for $H_{\rm{ext}}\perp (1,0,$\={2}$)$ . It is seen that,
 by taking into account the exchange anisotropy one can obtain an excellent agreement with the experimental data. This result is in quite contrast to the
  results from  Sirker \textit{et al.}  \ci{sirker} that were based on the HFP approximation alone.

 The optimized parameters $\gamma$ and U are universal for   both $H$$\perp$(1,0,\={2}) and
 $H$$\parallel$b cases. The main difference is only in the  g-factors. Using in the
  above equations
 $g=2.06$ we also  obtain the total magnetization for   $H$$\parallel$b, which is plotted
 in Fig.2(a) in comparison with the experimental data taken from \ci{nikinu}
 and with a corresponding calculaions based on the HFP approximations \ci{sirker}.
 It is seen that by neglecting the  anomalous density $\sigma$ (dashed line) or
 the exchange anisotropy , $\gamma$, (dotted line) one may reproduce the experimental magnetization only at high temperatures , $T>T_c$, while the inclusion
 of both, $\sigma$ and $\gamma$ makes it  possible to obtain a significantly  better theoretical description (solid line) for $T\leq T_c$ also. From the  Fig. 2(a) one may conclude that the effect of the exchange anisotropy is rather large in the BEC - like phase and is almost negligible  in the normal phase.

Another important characteristics of quantum magnets is that the magnetically ordered state
supports a staggered magnetization $M_\perp$ transverse to the field direction, leading to
a canted antiferromagnetic state until the system becomes eventually fully polarized
as the external field increases. In the BEC- scenario  the number of triplons corresponds to  the total magnetization $M_z$ along the field direction, while the number of condensed particles is proportional to the square of the ordered transverse component:
\be
M_\perp=N_f g\mu_B \sqrt{\rho_0/2},
\lab{mstag}
\ee
 where $N_f$ is a normalization factor \ci{nikinu}.  In our  present approximation
 $M_\perp$ may be calculated directly from Eqs. \re{zet1}-\re{rho0}   and \re{mstag}.
 The results are presented  in Fig.2 (b), where we have used the same input parameters
 as in Figs.1 and chosen the scaling factor $N_f=6.5$ to reproduce
  the experimental data \ci{tanaka}. As it is seen from the figure the present approach with exchange anisotropy
 describes well the experimental data for $T\leq T_c$. A comparison of  the  dotted curve with the solid  line  in Fig. 2(b) shows that the exchange anisotropy
 enhances the staggered magnetization.

 In the vicinity of the critical point the staggered magnetization scales as
 $M_\perp\sim (T_c-T)^\beta$, ($T\leq T_c$) defining the critical exponent $\beta$.
 Approximating the curves in Fig. 2(b) as
 \be
 M_\perp=\rm{const}\times(1-T/T_c)^\beta,
 \ee
 we have found $\beta=0.47$ in the present approximation, which is close to the
 predictions made in  Quantum Monte Carlo simulations  \ci{nohadani}: $\beta_{QMC}=1/2$.
 The other curves in Fig.2 (b) lead to $\beta=0.15$ and $\beta=0.39$ for HFP and
 HFB with $\gamma=0$  cases, respectively.

 In the present work we have been dealing  only with the  exchange anisotropy, which gives a
sharp phase transition with $\rho_0(T\geq T_c)=0$.
  Comparing our magnetization curves for the total magnetization
with the experimental data (see Fig.1(b) and Fig. 2(a)) we may conclude that including a finite exchange anisotropy is sufficient. However,  as it has been shown by Sirker \textit{et al.}
 the inclusion  of a
DM anisotropy instead may lead to a crossover \ci{sirker}, so that  $M_{\perp}^{\rm{DM}}(T\geq T_c)\neq 0$. Indeed,
  from the  fact that  experimental data
on the tranverse magnetization  show  $M_{\perp}^{\rm{exp}}(T\geq T_c)\neq 0$, (see Fig. 2(b) with data from Fig.3 in Ref. \ci{tanaka})  one may conclude that
a certain DM anisotropy is clearly present. Moreover, Density Matrix Renormalization Group
calculations \ci{mila} show that even a tiny DM interaction can modify some aspects
of the physics, especially the staggered magnetization, rather dramatically.
We shall develop a  HFB approximation  including both exchange and  DM anisotropies in a subsequent
 publication and do not discuss it here any further.

From Fig. 1(a) we can state  that the exchange anisotropy term $H'_{\rm{EA}}$
 given in \re{hamilt} leads to an increase of the critical temperature at a given magnetic
 field. To study this issue in  more detail
  we present in Fig. 3(a) the shift of the critical temperature due to the anisotropy $\Delta T/T_{c}^{0}$  vs $\gamma$
 for various values of $H_{\rm{ext}}$. We see that
 \begin{itemize}
 \item
 $\Delta T_c$ increases with the increase of $\gamma$;
 \item
 For a moderate value of gamma $\gamma\sim 0.04 \rm{K}$ the shift is nearly $10\%$
  at $ H_{\rm{ext}}=7\rm{T}$;
 \item
 With increasing  the magnetic field, the upward shift in the  critical temperature decreases.
 \end{itemize}
A similar  dependence of the shift on $\gamma$ and $H_{\rm{ext}}$ has been predicted by Dell'Amore
\textit{et al.} \ci{andreasaniz}.

 There is another   effect due to the explicit $U(1)$ - symmetry breaking.
  In real systems the presence of an  anisotropy modifies the energy dispersion
  of the magnetic excitations. Experimentally, the excitation
  spectrum of  TlCuCl$_3$ was investigated  by R\"{u}egg \textit{et al.} \ci{rueggnat}
  using  INS measurements.
   In the frame of
Bogolyubov mean field theory this spectrum coincides with spectrum of quasiparticles (called as bogolons) which can be calculated from equation \re{gap} in the present approximation. The energy dispersion of the low- lying magnetic excitations measured for
$H$$_{\rm{ext}}$=14T at temperatures $T$=50\rm{mK} and $T$=1.5 \rm{K} are presented in Fig.3(b), where the solid line
is obtained in the HFB approximation \re{gap} using our optimized parameters. It can be
 seen
that the agreement with the experiment is satisfactory, especially at small momentum transfer.
 Clearly in the BEC phase without anisotropy
the energy dispersion is linear at small momentum, i.e. $E_{k\rightarrow 0} (H_{\rm{ext}},T,\gamma=0)\sim ck$ , (with c is the  sound velocity),
 while the presence of an anisotropy
causes a gap  $\Delta E=E_{k\rightarrow 0} (H_{\rm{ext}},T,\gamma)\neq 0$ which
     can be calculated directly from Eq. \re{gap}.
     The average sound velocity at small momentum defined as
      $c=(\partial E/\partial k)\mid_{k\rightarrow 0}$ is given by
      $c(\gamma=0)=\sqrt{X_1/2m}$  at $T\leq T_c$ and $c(\gamma\neq 0)=k(X_1+X_2)/2m \sqrt{X_1 X_2}$ at any temperature, where $m$ is the effective mass
      \ci{misguich}
      and $X_1, X_2$ are given by Eq. \re{x12}.
     Fig. 3(c) illustrates the fact that,
 being zero at $\gamma=0$, the gap in the quasiparticle spectrum  increases with $\gamma$.
 In the present approach,  $\Delta E=E_{k\rightarrow 0} (H_{\rm{ext}}=14\rm{T}, T=1.5\rm{K},\gamma=0.045 \rm{K})=0.7 \rm{K}$ which is the detection limit of Ref. \ci{rueggnat}.

 A possible  modification of the spin gap separating the singlet ground state
 from the lowest - energy triplet excitation $\Delta_{\rm{st}}$ due the anisotropy,
 is not considered here, and we used the  experimental value $\Delta_{\rm{st}}=7.55 \rm{K}$
 \ci{misguich} (see appendix A).


\begin{figure}[H]
\begin{minipage}[H]{0.49\linewidth}
\center{\includegraphics[width=1.1\linewidth]{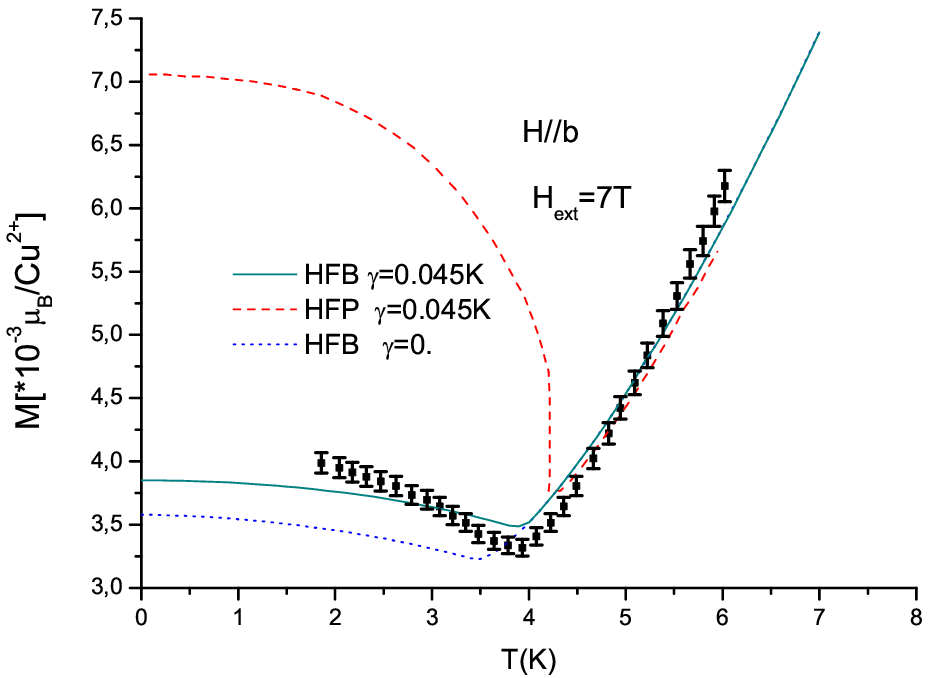} \\ a)}
\end{minipage}
\hfill
\begin{minipage}[H]{0.49\linewidth}
\center{\includegraphics[width=1.1\linewidth]{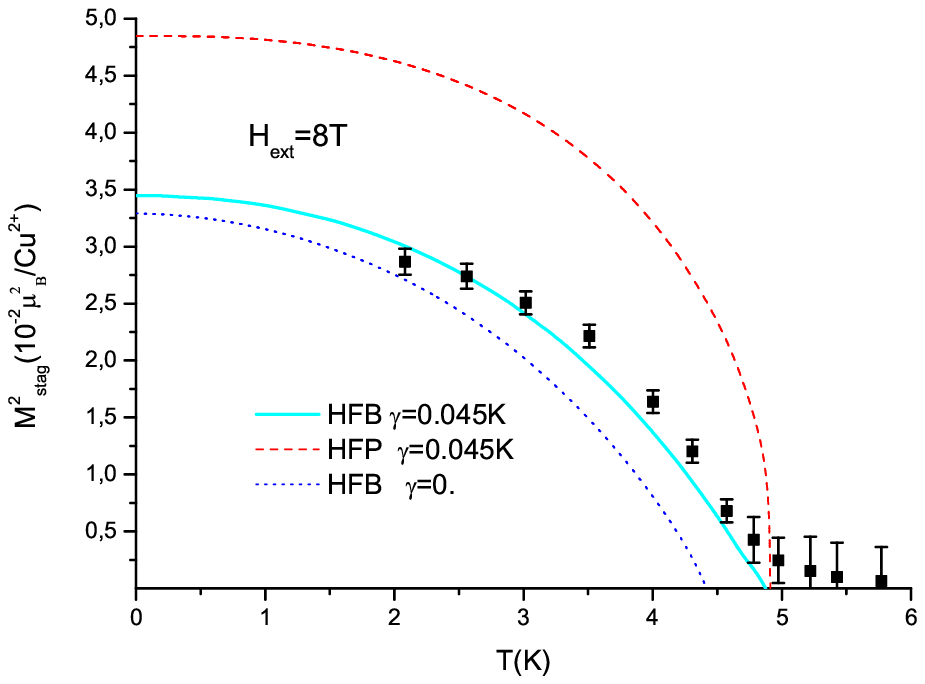} \\ b)}
\end{minipage}
\caption
{
The total magnetization (a)  and square of the transverse magnetization (b)  in different approximations  for $H\|b$. The solid and dashed lines are for HFB and HFP apppoximations with the
exchange anisotropy and the same input parameters, respectively. The dashed line
in (a) is reproduced from Ref. \ci{sirker} and the experimental data are taken
from \ci{nikinu,tanaka}. The dotted line represents a HFB approximation without anisotropy, i.e. $\gamma=0$, and $U=367.5 \rm{K}$
}
\label{fig2ab}
\end{figure}


 Finally we discuss the role of the anomalous density whose absolute value is the density
 of pair correlated particles. This pair correlations are, actually, responsible
 for the existence of superfluidity \ci{yukan}.
  We present in Fig.4(a) the density of condensed particles
  $\rho_0$ (solid line) and the absolute value of the  anomalous density
 $\mid\sigma\mid$ (dashed line) versus the reduced temperature.
 \footnote{Actually,  $\sigma<0$ in the whole region of temperatures.}
  It is seen that $\mid\sigma\mid$ is comparable with $\rho_0$ at all temperatures. Another interesting fact, which is demonstrated  in
   Fig. 4(b)   is that the anomalous density surives, although on a small level, even
 above the critical  temperature where it  vanishes asypmtotically. For example,
 $\sigma(t=0)/\sigma(t=1)\approx 100$.
  Note that without exchange anisotropy  $\sigma(\gamma=0)\mid_{T\geq T_c}=0$.
 A similar phase with $\rho_0=0$ and $\sigma\neq 0$ has been reported by Cooper \textit{et al.}
 \ci{cooper} within a lowest - order auxiliary field formalism. In fact, this approach predicts the existence of two critical temperatures, one $T_c$, where  $\rho_0=0$, $\sigma\neq0$ and another one  $T^*$, where  $\rho_0=0$, $\sigma=0$, with $T^*>T_c$. This exotic state in the region $T_c<T<T^*$ has not been experimentally observed yet, but it is predidicted to exhibit a modified dispersion relation. The question about the observing  such phase  still remains open.
\begin{figure}[H]
\begin{minipage}[H]{0.32\textwidth}
\center{\includegraphics[width=\linewidth]{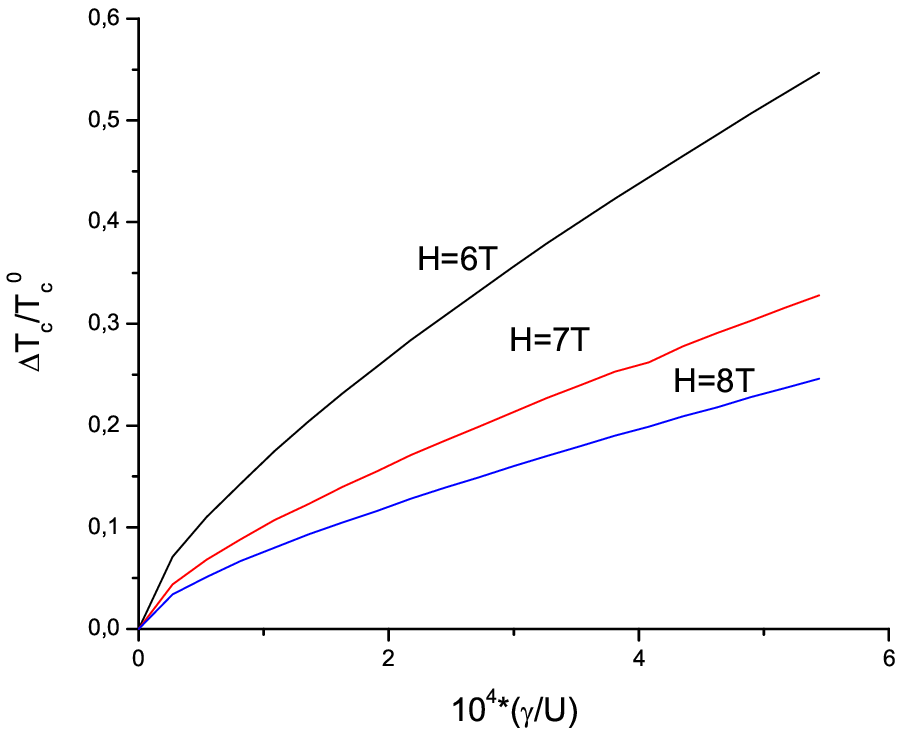} \\ a)}
\end{minipage}
\hfill
\begin{minipage}[H]{0.32\textwidth}
\center{\includegraphics[width=\linewidth]{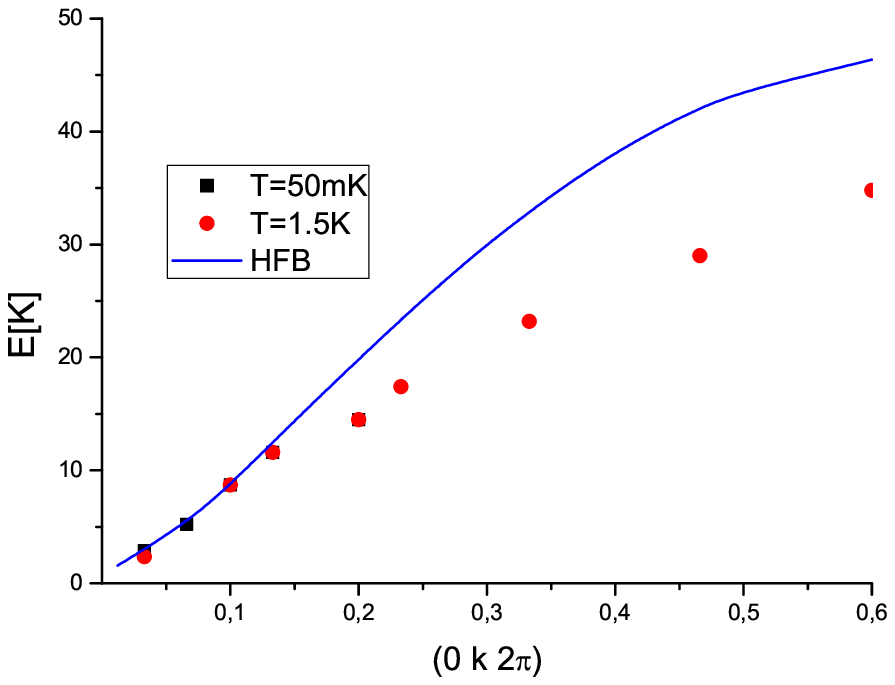} \\ b)}
\end{minipage}
\begin{minipage}[H]{0.32\textwidth}
\center{\includegraphics[width=\linewidth]{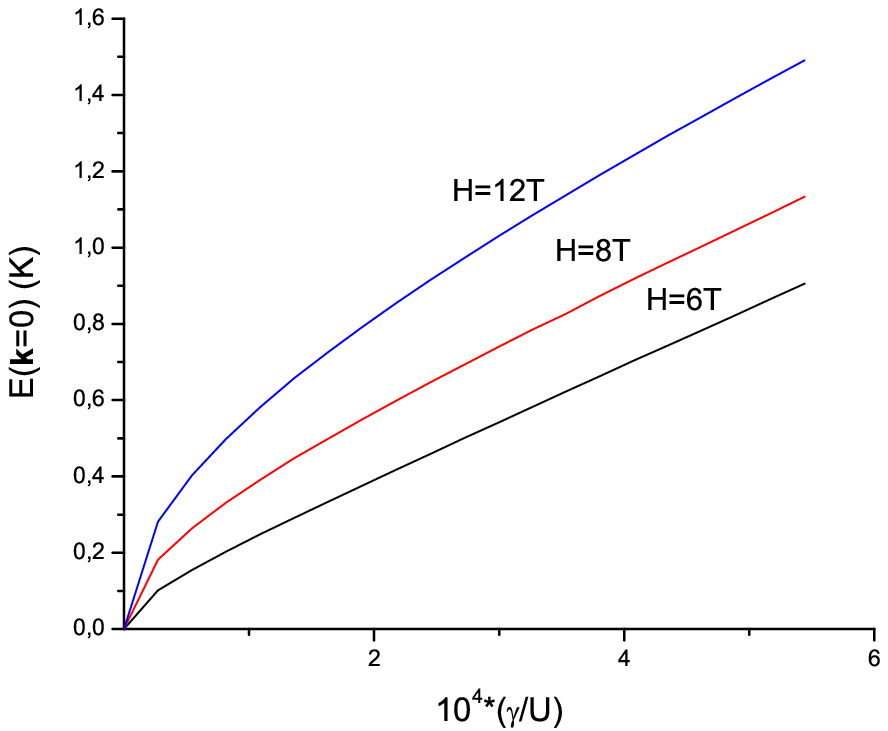} \\ c)}
\end{minipage}
\hfill
\caption{
 (a): The shift of the critical temperature,
$(T_c(\gamma)-T_c(\gamma=0))/T_c(\gamma=0)$
due to the anisotropy ;  (b): Quasiparticle
spectrum (solid line) \re{gap} in HFB approximation with the optimized parametrs for
$H=14 \rm{T}$ at $T\leq 1.5 \rm{K}$ . INS data for $T=50 \rm{mK}$ (circles) and $T=1.5 \rm{K}$ (squares) taken
from Ref. \ci{rueggnat} are shown for comparision. (c): The collective excitation gap $E_{k=0}$ in Eq. (\ref{gap})
in the BEC phase  vs.  $\gamma$ at $T$=1.5 \rm{K}.;.
 }
\label{fig3ab}.
\end{figure}

\begin{figure}[H]
\begin{minipage}[H]{0.49\linewidth}
\center{\includegraphics[width=1.1\linewidth]{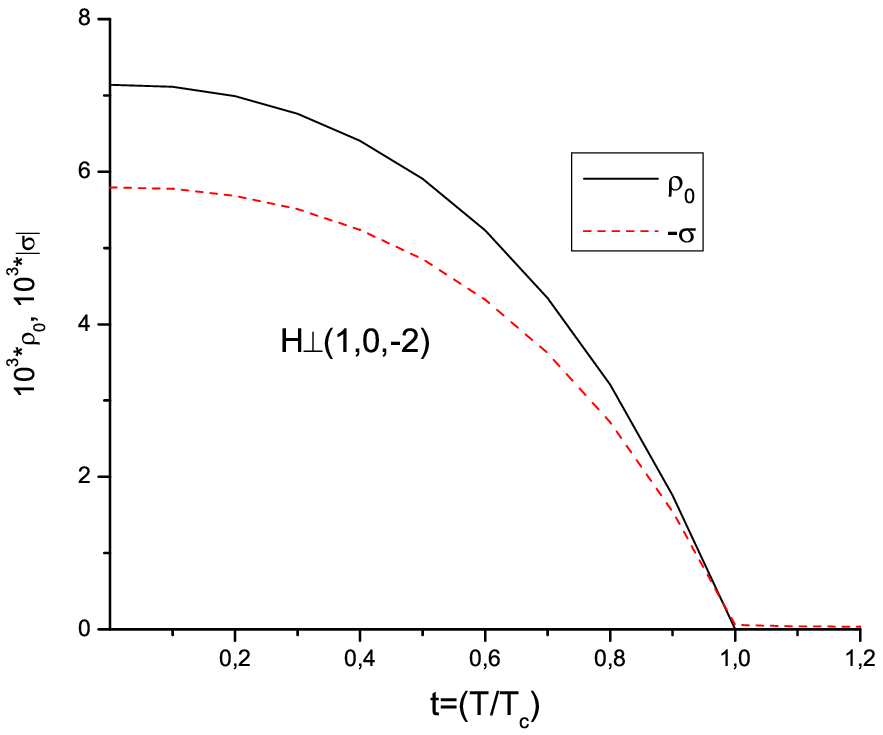} \\ a)}
\end{minipage}
\hfill
\begin{minipage}[H]{0.49\linewidth}
\center{\includegraphics[width=1.1\linewidth]{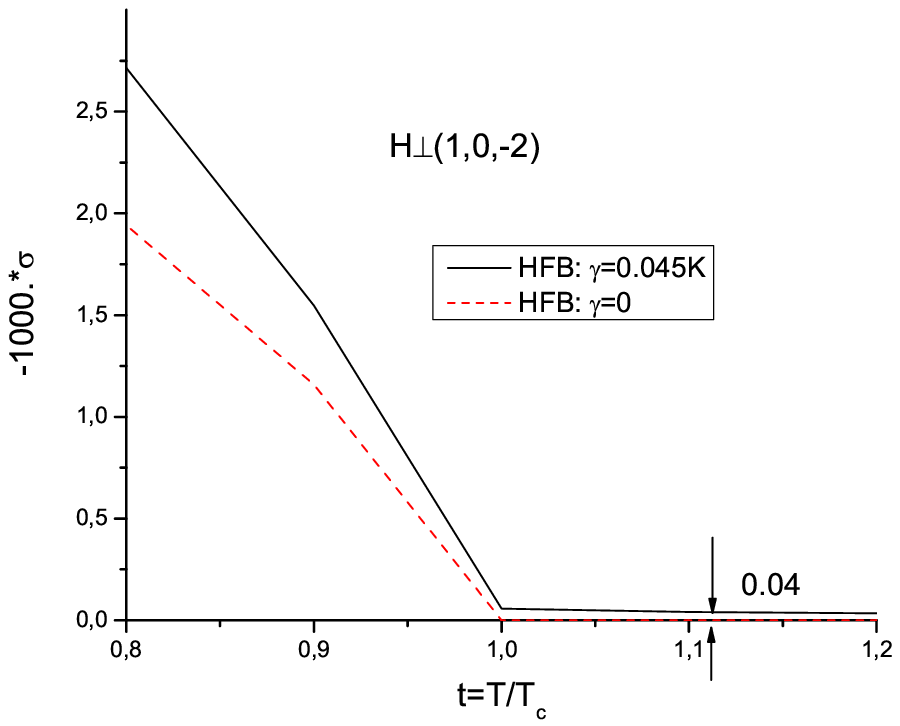} \\ b)}
\end{minipage}
\caption{
(a): The condensed (solid line) and the absolute value of anomalous densities
 (dashed line) in HFB approximation at $H_{\rm{ext}}=7 \rm{T}$; (b): The
   behavior of $\mid\sigma (t)\mid$
near the critical temperature
with (solid line) and without (dashed line) exchange anisotropy.
}
\label{fig4ab}.
\end{figure}


\section{Conclusion}

Assuming that  the low temperature properties of quantum magnets with a weak $U(1)$
symmetry breaking can be described in a BEC - like scenario, we proposed a new
MFA based  approach within the Hartree - Fock - Bogolyubov approximation, which takes into account an anomalous density $\sigma$ and exchange anisotropy.
This approach not only reproduces experimental data such as the critical temperature and the magnetization in a satisfactory way, but also removes certain inconsistencies and drawbacks met in the previous Hartree - Fock - Popov approaches \ci{sirker}.
%
 Remarkably, this may be reached by optimizing
only three paramters : $\gamma$- the parameter of the exchange anisotropy, $c_{\gamma}$- the parameter of breaking of the Hugenholtz -- Pines relation and U - the interparticle interaction. We have found $\gamma=0.045 \rm{K}=0.0038 \rm{meV}$, $c_{\gamma}=1.67$ and $U=367.5 \rm{K}$ valid
for both  $H{\parallel} b$ and $H\perp (1,0,\bar{2})$ directions.

The present approach also  gives a fair  theoretical description of the staggered
magnetization data for $T\leq T_c$ and predicts a plausible value for the critical exponent
$\beta$. However, to improve the theoretical description
of the experimental data on staggered magnetization the inclusion of
Dzyaloshinsky-Moriya anisotropy also seems to be nessesary.
 We have estimated the anisotropy - induced  shift of the
critical temperature, and show that it is substantial.
Finally, we predict that the anomalous density is comparable to the condensded one, and survives at temperatures exceeding $T_c$ where the condensate fraction is zero.
\section*{Acknowledgments}

We are indebted to E. Ya. Sherman and V.I. Yukalov for useful discussions and comments.  This work is partially supported by the Swiss National Foundation SCOPES project
$IZ74Z0\_160527$.


\newpage
\section*{Appendix A }
\def\theequation{A.\arabic{equation}}
\setcounter{equation}{0}

The summation by momentums may be explicitly written as
\bea
\fl
\sum_{\mathbf{k}}f(\varepsilon(k))=\frac{1}{{(2\pi)^{3}}}\left\{
  \begin{array}{ll}
    {4\pi}\int\limits_{0}^{\infty}k^{2}dk f(\varepsilon(k)),    \quad \varepsilon(k)=\frac{k^{2}}{2m}  \quad, & \hbox{} \\
    \int\limits_{-\pi}^{\pi}dk_{x}dk_{y}dk_{z}f(\varepsilon(k)),   \quad \varepsilon(k)-\textrm{isotropic} \quad  ,& \hbox{} \\
    \frac{1}{2}\int\limits_{-\pi}^{\pi}dk_{x}\int\limits_{-\pi}^{\pi}dk_{y}\int\limits_{-2\pi}^{2\pi}dk_{z}\varepsilon(k), \quad  (\varepsilon(k))-\textrm{realistic, anisotropic}  . & \hbox{}
  \end{array}
\right.
\eea
The isotropic bare dispersion may be presented as $\varepsilon(k)=J[3-\cos k_{x}a-\cos k_{y}a-\cos k_{z}a]$
where $a$ is the size of the unit cell ( below we set $a=1$), while the anisotropic one may be written as \ci{misguich}
\be
\ba
\varepsilon_{k-k_{0}}=-\Delta_{st}+\sqrt{(J+\tilde{a})^{2}-\tilde{a}^2} ,\\
\tilde{a}=J_{a}\cos(k_{x})+J_{a2c}\cos(2k_{x}+k_{z})+2J_{abc}
\cos(k_{x}+k_{z}/2)\cos(k_{y}/2).
\ea
\ee
In practical calculations with this realistic dispersion one may make a shift as $\mathbf{k}-\mathbf{k}_{0}\rightarrow \mathbf{k}$,  so that $\varepsilon(k-k_{0})|_{k=k_{0}}\rightarrow\varepsilon(k)|_{k=0}=0$, $k_{0}=\{0,0,2\pi\}$   and introducing $q_{x}=k_{x}/\pi$, $q_{y}=k_{y}/\pi$, $q_{z}=k_{z}/4\pi$ we can rewrite the summation as
\bea
\sum_{\mathbf{k}}f(\varepsilon(k))|_{aniz}=\frac{1}{2}
\int\limits_{-1}^{1}dq_{x}\int\limits_{0}^{1}dq_{y}\int\limits_{0}^{1}dq_{z}f(\varepsilon(q)),  \eea
where    $\varepsilon_{q}=-\Delta_{st}+\sqrt{J^{2}+2Ja_{q}}$, and
\bea \label{aq}
\fl
a_{q}=J_{a}\cos(\pi q_{x})+J_{a2c}\cos(2\pi q_{x}-4\pi q_{z})
-2J_{abc}\cos(\pi q_{x}-2\pi q_{z})\cos(\pi q_{y}/2) .
\eea
The condition $\veps_q (q=0)=0$ fixes $\Delta_{st}$ as $\Delta_{st}=
\sqrt{J^2+2J(J_a+J_{a2c}-2J_{abc})}$.
In the present work we used the  following values of parameters \ci{misguich}:
$J=63.7$K, $J_a=-2.5K$, $J_{a2c}=-18.35$K and $J_{abc}=5.28K$, so that $\Delta_{st}=7.55K$.


\section*{Appendix B }
\def\theequation{B.\arabic{equation}}
\setcounter{equation}{0}

In the notion of representative ensemble \ci{yukan} the grand Hamiltonian including the exchange anisotropy term can be  written as:
\be
\ba
\fl
H=\hat{H}-\mu_{0}\hat{N}_{0}-\mu_{1}\hat{N}_{1}-\hat{\Lambda},\\
\fl
\hat{H}=\int\left\{\psi^{\dag}(\mathbf{r})\hat{K}\psi(\mathbf{r})+\frac{U}{2}\left(\psi^{\dag}(\mathbf{r})\psi(\mathbf{r})\right)^{2}
+\frac{{\gamma}}{2}\left(\psi^{\dag}(\mathbf{r})\psi^{\dag}(\mathbf{r})+\psi(\mathbf{r})\psi(\mathbf{r})\right)\right\}d^3r ,
\lab{grandH}
\ea
\ee
where $\hat{N}_{0}=\int\mid\phi_0\mid^{2}d^3r$, \quad $\hat{N}_{1}=\int\widetilde{\psi}^{\dag}(\mathbf{r})\widetilde{\psi}(\mathbf{r})d^3r$, \quad
so that  $\mu N=\mu_{0}N_{0}-\mu_{1}N_{1}$,  $\hat{N}=\int\psi^{\dag}\psi d^3r$  \qquad
is the total number of particles.
The Lagrange multiplier
  \be
\hat{\Lambda}=\int[\lambda\widetilde{\psi}^{\dag}(\mathbf{r})+\lambda^{\dag}\widetilde{\psi}(\mathbf{r})]d^3r
\lab{lambda}
\ee is a so called  linear killer, such that $\lambda$ is chosen from the constraint for the conservation of quantum numbers,
$\langle \widetilde{\psi}(\mathbf{r})\rangle=0$. The quantum fluctuation $\widetilde{\psi}(\mathbf{r})$ is related to the field
operator as $\psi(\mathbf{r})=\phi_0+\widetilde{\psi}(\mathbf{r})$, which makes possible to rewrite
the grand Hamiltonian as follows:
\begin{equation}
H=H_{0}+H_{1}+H_{2}+H_{3}+H_{4} ,
\end{equation}
where
\be
H_{0}=\int d^3r \rho_0(-\mu_{0}+\gamma+\frac{U}{2}\rho_{0})
\lab{H0}
\ee
with $\rho_{0}=\phi_{0}^{2}$
\be
\fl
H_{1}=\int d^3r\phi_0 (\gamma+U\rho_0)(\widetilde{\psi}^{\dag}(\mathbf{r})+\widetilde{\psi}(\mathbf{r})),
\lab{H1}
\ee
\begin{equation}
\fl
H_{2}=\int d^3r\left\{\widetilde{\psi}^{\dag}(\mathbf{r})[\hat{K}-\mu_{1}+2U\rho_{0}]\widetilde{\psi}(\mathbf{r})
+\frac{1}{2}\left(\widetilde{\psi}^{2}(\mathbf{r})+\widetilde{\psi}^{\dag2}(\mathbf{r})\right)
({\gamma}+U\rho_{0})\right\},
\lab{H2}
\end{equation}
\be
\fl
H_{3}=U\phi_0\int d^3r \widetilde{\psi}^{\dag}(\mathbf{r})\widetilde{\psi}(\mathbf{r})\left(\widetilde{\psi}^{\dag}(\mathbf{r})+\widetilde{\psi}(\mathbf{r})\right) ,
\lab{H3}
\ee
\begin{equation}
\fl
H_{4}=\frac{U}{2}\int d^3r(\widetilde{\psi}^{\dag}(\mathbf{r})\widetilde{\psi}(\mathbf{r}))^{2} .
\lab{H4}
\end{equation}
Performing a Fourier transformation $\widetilde{\psi}(\mathbf{r})=\sum_{\mathbf{k}\neq 0} a_\mathbf{k} \widetilde{\psi}_\mathbf{k}(\mathbf{r}) $
and assuming that $\phi_0(\mathbf{r})$ does not depend on ${\mathbf{r}}$
 we may rewrite the  above equations as follows:
 \be
 \ba
 H_{0}=\rho_0(-\mu_{0}+\gamma+\frac{U}{2}\rho_{0}),\\
 H_1=\sqrt{\rho_0}(U\rho_0+\gamma)\sum_{\mathbf{k}}(a_{\mathbf{k}}+a_{\mathbf{k}}^{\dag}),\\
H_{2}=\sum_{\mathbf{k}}\{\varepsilon_{k}-\mu_{1}+2U\rho_{0}\}
a_{\mathbf{k}}^{\dag}a_{\mathbf{k}}+\frac{1}{2}(a_{\mathbf{k}}a_{-\mathbf{k}}+a_{\mathbf{k}}^{\dag}a_{-\mathbf{k}}^{\dag})
({U\rho_0+\gamma}),\\
H_3=U\sqrt{\rho_0}\sum_{\mathbf{k},\mathbf{p}}(
a_{\mathbf{k}}^{\dag}a_{\mathbf{k}+\mathbf{p}}a_{-\mathbf{p}}+\rm{h.c.}
),\\
H_{4}=\displaystyle{\frac{U}{2}}\sum_{\mathbf{k},\mathbf{q},\mathbf{p}}a_{\mathbf{k}}^{\dag}a_{\mathbf{p}}^{\dag}a_{\mathbf{p}+\mathbf{q}}a_{\mathbf{k}-\mathbf{q}} .
\ea
\ee
The third and fourth order terms may be further simplified by using
the  approximation given in Eq. \re{wick} as:
\be
\ba
H_3=U\sqrt{\rho_0}(2\rho_1+\sigma)\sum_{\mathbf{k}}(a_{\mathbf{k}}^{\dag}+a_{\mathbf{k}}),\\
H_{4}=\displaystyle{\frac{U}{2}}\{4\rho_{1}a_{\mathbf{k}}^{\dag}a_{\mathbf{k}}+\sigma(a_{\mathbf{k}}a_{-\mathbf{k}}
+a_{-\mathbf{k}}^{\dag}a_{\mathbf{k}}^{\dag})-(2\rho_{1}^{2}+{\sigma}^{2})\} .
\lab{h34}
\ea
\ee

Now the grand Hamiltonian is the sum of classical, $H_{\rm{class}}$, linear $H_{\rm{lin}}$ and
$H_{\rm{bilin}}$ terms as:
\be
\ba
H=H_{\rm{class}}+H_{\rm{lin}}+H_{\rm{bilin}},\\
H_{\rm{class}}=-\rho_0\mu_{0}+\rho_0\gamma+\displaystyle{\frac{U}{2}}\rho_{0}^{2}
-\displaystyle{\frac{U}{2}}(2\rho_{1}^{2}+{\sigma}^{2}),\\
H_{\rm{lin}}=U\sqrt{\rho_0}(\rho_0+\widetilde{\gamma}+2\rho_1+\sigma)\sum_{\mathbf{k}}(a_{\mathbf{k}}^{\dag}+a_{\mathbf{k}}),\\
H_{\rm{bilin}}=\sum_{\mathbf{k}}(\varepsilon_{k}-\mu_{1}+2U\rho)
a_{\mathbf{k}}^{\dag}a_{\mathbf{k}}+
\displaystyle{\frac{U({\rho_0+\widetilde\gamma+\sigma})}{2}}\sum_\mathbf{k}(a_{\mathbf{k}}a_{-\mathbf{k}}+a_{\mathbf{k}}^{\dag}a_{-\mathbf{k}}^{\dag}),
\lab{linbilin}
\ea
\ee
where $\widetilde\gamma=\gamma/U$ and $\rho=\rho_0+\rho_1$.
In the formalism of representative ensemble \ci{yukan} the linear term is neglected
by an appropriate choice of $\lambda$, for example, by choosing $\lambda=U\sqrt{\rho_0}(\rho_0+\widetilde{\gamma}+2\rho_1+\sigma)$ in Eq. \re{lambda}.
The $\mu_0$  can be found by minimization of the free energy with respect
to $\rho_0$.
To diagonalize the bilinear term we introduce the  normal $\Sigma_n$ and
anomalous $\Sigma_{an}$ self energies as
\be
\ba
\Sigma_n=2U\rho,\\
\Sigma_{an}=U(\rho_0+\widetilde\gamma+\sigma) ,
\lab{self}
\ea
\ee
such that $H_{\rm{bilin}}$ is rewritten as
\be
H_{\rm{bilin}}=\sum_{\mathbf{k}}\omega_k
a_{\mathbf{k}}^{\dag}a_{\mathbf{k}}+
\frac{\Sigma_{an}}{2}\sum_\mathbf{k}(a_{\mathbf{k}}a_{-\mathbf{k}}+a_{\mathbf{k}}^{\dag}a_{-\mathbf{k}}^{\dag}),
\lab{bilin}
\ee
where
\be
\omega_k=\varepsilon_{k}-\mu_{1}+\Sigma_n .
\ee

 The next step is the Bogolyubov transformation
\bea
a_{\mathbf k}=u_{\mathbf k}b_{\mathbf k}+v_{\mathbf k}b_{-{\mathbf k}}^{\dagger}, \quad
a_{\mathbf k}^{\dagger}=u_{\mathbf k}b_{\mathbf k}^{\dagger}+v_{\mathbf k}b_{-{\mathbf k}}
\lab{abog}
\eea
to diagonalize ${H}_{\rm{bilin}}$.
The operators $b_{\mathbf k}$ and $b_{\mathbf k}^{\dagger}$ can be interpreted as annihilation and creation operators of phonons with following properties:
\begin{eqnarray}
&&[b_{\mathbf k},b_{\mathbf p}^{\dagger}]=\delta_{\mathbf{k},\mathbf{p}},\quad  \langle b_{\mathbf k}^{\dagger}b_{-{\mathbf k}}^{\dagger}\rangle=\langle b_{\mathbf k}b_{-\mathbf{k}}\rangle=0,\\
&&\langle b_{\mathbf k}^{\dagger}b_{\mathbf k}\rangle=f_{B}(E_{k})=\dsfrac{1}{e^{\beta E_{k}}-1},
\lab{2.23}
\end{eqnarray}
where $\beta\equiv 1/T$. To determine the phonon dispersion $E_{k}$ we insert \re{abog} into \re{bilin} and require that the coefficient
of the term $b_{\mathbf k}b_{-\mathbf{k}}+b_{-\mathbf{k}}^{\dagger}b_{\mathbf k}^{\dagger}$ vanishes, i.e:
\bea
\omega_{k}u_{\mathbf k}v_{\mathbf k}+\frac{\Sigma_{an}}{2}\left(u^{2}_{\mathbf k}+v_{\mathbf k}^{2}\right)=0.
\lab{2.24}
\eea
Now using the condition $u_{\mathbf k}^{2}-v_{\mathbf k}^{2}=1$ and presenting $u_{\mathbf k}, v_{\mathbf k}$ as
\bea
u_{\mathbf k}^{2}=\frac{\omega_{k}+E_{k}}{2E_{k}},\quad \quad v_{\mathbf k}^{2}=\frac{\omega_{k}-E_{k}}{2E_{k}}
\lab{2.25}
\eea
yields
\bea
\sqrt{\omega_{k}^{2}-E_{k}^{2}}=-\Sigma_{an},\quad\quad u_{\mathbf k}v_{\mathbf k}=-\frac{\Sigma_{an}}{2E_k},\quad\quad u_{\mathbf k}^2+v_{\mathbf k}^2=\frac{\omega_k}{E_k}
\eea
that is
\be
\ba
E_k^2=(\omega_k+\Sigma_{an})(\omega_k-\Sigma_{an})
\equiv(\varepsilon_k+X_1)(\varepsilon_k+X_2)
\lab{Ekk}
\ea
\ee
with
\bea
\left\{
  \begin{array}{ll}
    X_{1}=\Sigma_n+\Sigma_{an}-\mu_1 & \hbox{} \\
    X_{2}=\Sigma_n-\Sigma_{an}-\mu_1 & \hbox{}.
    \lab{xx12}
  \end{array}
\right.
\eea
Now $H_{\rm{bilin}}$ is simplified as
\be
H_{\rm{bilin}}=\sum_\mathbf{k}E_kb_\mathbf{k}^{\dag}b_\mathbf{k}+\frac{1}{2}\sum_\mathbf{k}(E_k-\omega_k)
\ee
and the total Hamiltonian is given by

\be
\fl
H=H_{\rm{class}}+H_{\rm{bilin}}=-\rho_0\mu_{0}+\rho_0\gamma+\frac{U}{2}\rho_{0}^{2}
-\frac{U}{2}(2\rho_{1}^{2}+{\sigma}^{2})+
\sum_\mathbf{k}E_kb_\mathbf{k}^{\dag}b_\mathbf{k}+\frac{1}{2}\sum_\mathbf{k}(E_k-\omega_k)
\ee
which may be used to define the energy of the system.

Note that by requiring in \re{Ekk} $X_2=0$, one may directly obtain from Eqs.
\re{xx12} the
Hugenholtz - Pines theorem as well as the gapless dispersion in SSB phase.
The main Eqs. \re{x1eta1} and \re{x2eta2} are derived by inserting \re{xx12} into \re{self}.
 The normal and anomalous densities  may be obtained by using Eqs. \re{abog}
in Eqs. \re{rho11} and \re{sigma1} leading to the expressions \re{ro} and \re{sigm}
respectively, where $X_1$ and $X_2$ are given in \re{xx12} and \re{self}.

\section*{References}
\begin{thebibliography}{99}

\bibitem{cheng}  Cheng T.P. and   Li L.F. 1988 {\it
 Gauge theory of elementary particle physics} (Oxford University Press).
\bibitem{ginibre} Ginibre J. {\it Commun. Math.Phys.} 1968 {\bf 8} 26
\bibitem{oosawa}  Oosawa A.  et al. 1999 {\it J. Phys. Condens. Matter} {\bf 11}  265
 \bibitem{rueggnat}  R\"{u}egg Ch. et al. 2003 {\it Nature (London)} {\bf 423} 62
 \bibitem{yamada}  Yamada F. et al. 2008 {\it J. Phys. Soc. Jpn.} {\bf 77} 013701
 \bibitem{nikinu}  Nikuni T. et al. 2000 {\it Phys. Rev. Lett.}  {\bf 84} 5868
 \bibitem{yuktriplon} V. Yukalov 2012 {\it Laser Physics} {\bf 22} 1145
\bibitem{zapf}  Zapf V. and  Jaime M.  2014 {\it Rev.Mod.Phys} {\bf 86} 563
\bibitem{cizmar}  \v{C}i\v{z}m\'{a}r E. et al. 2010 {\it  Phys Rev.} {\bf B 82} 054431
 \bibitem {rueggprl}  R\"{u}egg Ch. et al. 2008 {\it  Phys. Rev. Lett.} {\bf 100} 205701
 \bibitem{giamarchi}  Giamarchi T.  et al. 2008 {\it Nat. Phys.} {\bf 4} 198
 \bibitem{andreasaniz}  Dell'Amore R.,  Schilling A. and
  Kr\"{a}mer K.  2009 {\it Phys. Rev.} {\bf B
  79} 014438
 \bibitem{sirker}  Sirker J.,   Weisse A., and  Sushkov O. P. 2004 {\it Europhys. Lett.}
 {\bf 68} 275
\bibitem{ourphysrev}  Rakhimov A. ,  Sherman E. Ya., and  Kim Chul Koo
2010 {\it Phys. Rev.}
{\bf B 81} 020407(R)
\bibitem{ourannals}  Rakhimov A. ,  Mardonov S. , and  Sherman E. Ya.
2011 {\it Ann.  Phys.}
{\bf 326} 2499
\bibitem{dodds}  Dodds T.  Yang B.,
and  Kim Y. 2010 {\it Phys. Rev.} {\bf  B 81} 054412
\bibitem{hohenberg} Hohenberg P.C. and Martin P. 1965 {\it Ann.  Phys.} {\bf 34} 291
\bibitem{pines}  Hugenholtz N. M.  and  Pines D.
 1959 {\it Phys. Rev.} {\bf 116} 489
\bibitem{yukkl}  Yukalov V. I. and   Kleinert H. 2006
 {\it Phys. Rev.} {\bf A 73} 063612
 \bibitem{ournjp}  Rakhimov A. et al. 2012 {\it
 New J. Phys. } {\bf 14}  113010
 \bibitem{yukan}  Yukalov V. I. 2011 {\it Physics of Particles and Nuclei}  {\bf 42} 460
\bibitem{Kleinert05}     Kleinert H.,  Schmidt S. , and  Pelster A. 2005
                   {\it Annalen der Physik (Leipzig)}  {\bf 14} 214 .
\bibitem{AndersenRevModPhys}  Andersen J. 2004
                     {\it Rev. Mod. Phys.} {\bf 76} 599
\bibitem{yukanom}  Yukalov V.I. and  Yukalova E.P. 2005 {\it Laser Phys. Lett.} {\bf  2} 506
\bibitem{Yukalov11}  Yukalov V. I.,  Rakhimov A. , and
 Mardonov S. 2011 {\it  Laser Physics}
{\bf  2} 264
\bibitem{stoofbook}  Stoof H. T. C.,  Gubbels K. B. and  Dickerscheid D.B.M. 2009
{\it Ultracold Quantum Fields} (Springer)
\bibitem{ourlat} Kleinert H.,  Narzikulov Z. and
  Rakhimov A. 2012 {\it Phys. Rev.} {\bf A 85} 063602
\bibitem{griffin}  Griffin A. 1996 {\it Phys. Rev.} {\bf B 53} 9341
\bibitem{delamore} Dell'Amore R.,  Schilling A.  and  Kr$\ddot{a}$mer K.
2008
{\it Phys. Rev.} {\bf B 78} 224403
\bibitem{misguich} Misguich  G. and  Oshikawa M. 2004 {\it J. Phys. Soc. Jpn.} {\bf 73} 3429
\bibitem{shermanprl}  Sherman E.Ya. et al. 2003
 {\it Phys. Rev. Lett.}  {\bf 91}  057201  .
\bibitem{yamadaglass}  Yamada F. 2011 et al. {\it Phys. Rev.} {\bf B 83} 020409(R)
\bibitem{kastening} Kastening B. 2003 {\it Phys. Rev.}  {\bf A68} 061601
\bibitem{tanaka} Tanaka H. et al. 2001 {\it J.  Phys.  Soc.  Jpn.
} {\bf 70}  939
\bibitem{nohadani}  Nohadani O.,  Wessel S.   and  Haas S. 2005
 {\it Phys. Rev.}
{\bf  B 72} 024440
\bibitem{mila}  Miyahara S. et al. 2007 {\it Phys. Rev.} {\bf B 75} 184402
\bibitem{cooper} Cooper  F. 2011 et al. {\it Phys. Rev.} {\bf A 83} 053622
\eb
\end{document}